\documentclass[]{aa}
\usepackage[]{graphicx}  
\usepackage{natbib}
\usepackage{txfonts}
\usepackage{color}

\usepackage{ulem}  
\usepackage[figuresright]{rotating}  

\def\pow#1#2{#1$\times$10$^{#2}$}
\def\powm#1#2{#1\times10^{#2}}
\def\micron{$\mu$m}
\def\kms{$\mathrm{km}\,\mathrm{s}^{-1}$}  
\def\Kkms{$\mathrm{K\,km}\,\mathrm{s}^{-1}$}  
\def\psqcm{$\mathrm{cm}^{-2}$}
\def\pccm{$\mathrm{cm}^{-3}$}
\def\Tmb{$T_\mathrm{mb}$}  
\def\Tastar{${T_A}^*$} 
\def\Taprime{${T_A}^\prime$} 
\def\Tkin{$T_\mathrm{kin}$}   
\def\etamb{$\eta_\mathrm{mb}$}  
\def\intintens{$\int T_\mathrm{mb} \mathrm{d}V$}
\def\vlsr{$V_\mathrm{lsr}$}  
\def\vcentr{$V_\mathrm{centr}$}  
\def\Eup{$E_\mathrm{up}$}  
\def\nHH{$n_\mathrm{H_2}$}  

\def\HH{H$_2$}
\def\HCOplus{HCO$^+$}

\def\HHCO{H$_2$CO}

\def\thCO{$^{13}$CO}
\def\CstO{C$^{17}$O}
\def\CeiO{C$^{18}$O}
\def\SOtwo{SO$_2$}

\def\methanol{CH$_3$OH}

\def\water{H$_2$O}

\definecolor{orange}{rgb}{0.8,0.4,0.0}
\definecolor{darkblue}{rgb}{0.0,0.0,0.6}

\def\bigbr{\\ \noalign{\smallskip}}

\def\markchanges{yes}  

\def\marked{yes}
\def\unmarked{no}

\definecolor{orange}{rgb}{1.0,0.15,0}
\definecolor{darkgreen}{rgb}{0.0,0.0,0.4}

\ifx\markchanges\marked 
	\newcommand{\removed}[1]{\textcolor{orange}{[}\sout{#1}\textcolor{orange}{]}}  
	
\fi
\ifx\markchanges\unmarked 
	\newcommand{\removed}[1]{}  
	
\fi

\defcitealias{kazmierczak_inprep}{Paper II}  

\begin{document}

\title{
The HIFI spectral survey of AFGL2591 (CHESS). I.~Highly excited linear rotor molecules in the high-mass protostellar envelope\thanks{{\it Herschel} is an ESA space observatory with science instruments provided by European-led Principal Investigator consortia and with important participation from NASA.}
}
\titlerunning{HIFI spectral survey of AFGL2591 (CHESS). I.~Highly excited linear rotor molecules}

\author{
M.~H.~D.~van der Wiel\inst{\ref{kapteyn},\ref{sron},\ref{isis}} 
\and 
L.~Pagani \inst{\ref{lerma}} 
\and
F.~F.~S.~van der Tak\inst{\ref{sron},\ref{kapteyn}} 
\and
M.~Ka{\'z}mierczak\inst{\ref{sron}}
\and
C.~Ceccarelli \inst{\ref{laog}} 
}

\institute{
Kapteyn Astronomical Institute, University of Groningen, P.O.~Box 800, 9700\,AV, Groningen, The Netherlands\\email: \texttt{matthijs.vanderwiel@uleth.ca}
\label{kapteyn}
\and SRON Netherlands Institute for Space Research, P.O. Box 800, 9700\,AV, Groningen, The Netherlands
\label{sron}
\and Institute for Space Imaging Science, Department of Physics \& Astronomy, University of Lethbridge, Lethbridge AB, T1K\,3M4, Canada
\label{isis}
\and LERMA, UMR8112 du CNRS, Observatoire de Paris, 61, Av.~de l\arcmin Observatoire, 75014 Paris, France
\label{lerma}
\and Laboratoire d'Astrophysique de Grenoble, UMR 5571-CNRS, Universit\'e Joseph Fourier, Grenoble, France
\label{laog}
}

\date{\today}

\abstract
{
Linear rotor molecules such as CO, \HCOplus\ and HCN are important probes of star-forming gas. For these species, temperatures of $\lesssim$50\,K are sufficient to produce emission lines that are observable from the ground at (sub)millimeter wavelengths. Molecular gas in the environment of massive protostellar objects, however, is known to reach temperatures of several hundred~K. To probe this, space-based far-infrared observations are required. 
}
{We aim to reveal the gas energetics in the circumstellar environment of the prototypical high-mass protostellar object AFGL2591. }
{Rotational spectral line signatures of CO species, \HCOplus, CS, HCN and HNC from a 490--1240\,GHz survey with {\it Herschel}/HIFI, complemented by ground-based JCMT and IRAM~30m spectra, cover transitions in the energy range (\Eup/$k$) between 5\,K and $\sim$300\,K. Selected frequency settings in the highest frequency HIFI bands (up to 1850\,GHz) extend this range to 750\,K for $^{12}$C$^{16}$O. The resolved spectral line profiles are used to separate and study various kinematic components. Observed line intensities are compared with a numerical model that calculates excitation balance and radiative transfer based on spherical geometry.}
{
The line profiles show two emission components, the widest and bluest of which is attributed to an approaching outflow and the other to the envelope. 
We find evidence for progressively more redshifted and wider line profiles from the envelope gas with increasing energy level. This trend is qualitatively explained by residual outflow contribution picked up in the systematically decreasing beam size. Integrated line intensities for each species decrease as \Eup/$k$ increases from $\lesssim$50 to $\sim$700\,K. 
The \HH\ density and temperature of the outflow gas are constrained to $\sim$$10^5$--$10^6$\,\pccm\ and 60--200\,K. In addition, we derive a temperature between 9 and 17\,K and $N$(\HH) $\sim$ \pow{3}{21}\,\psqcm\ for a known foreground cloud seen in absorption, and $N$(\HH) $\lesssim$ $10^{19}$\,\psqcm\ for a second foreground component. 
 }
{
Our spherical envelope model systematically underproduces observed line emission at \Eup/$k$ $\gtrsim$ 150\,K for all species. This indicates that warm gas should be added to the model and that the model's geometry should provide low optical depth pathways for line emission from this warm gas to escape, for example in the form of UV heated outflow cavity walls viewed at a favorable inclination angle. 
Physical and chemical conditions derived for the outflow gas are similar to those in the protostellar envelope, possibly indicating that the modest velocity ($\lesssim$10\,\kms) outflow component consists of recently swept-up gas. 
}


\maketitle

\section{Introduction}
\label{sec:2591highexcintro}

Because young high-mass stars ($\gtrsim 8\,M_\odot$) reach the hydrogen-burning phase \emph{before} the accretion phase ends \citep{palla1993}, their formation process differs from less massive protostellar objects. Several theories exist for the formation of massive stars \citep[see review by][]{zinnecker2007}, including stellar mergers and so-called `competitive accretion' \citep{bonnell2001a,bonnellbate2005}. However, the accumulating observational evidence for collimated outflows and disk structures in massive star-forming regions \citep{beuther2002b,cesaroni2005,sandell2010,kraus2010,wang2012} supports a picture of anisotropic accretion through a disk structure, analogous to the low-mass star formation process \citep[e.g.,][]{shu1987,larson2003}, at least for luminosities up to $10^5$\,$L_\odot$. Even with the latter view of high-mass star formation, striking differences arise compared to low-mass star formation theory: shorter timescales by a factor $\sim$$10$ and higher internal luminosities by several orders of magnitude. Moreover, the bulk of the energy produced by massive stars is emitted at UV wavelengths, by which the surrounding gaseous material is heated, partly ionized, and otherwise chemically influenced. The effect of UV radiation on circumstellar gas has been recognized for low-mass protostars \citep[e.g.,][]{spaans1995} and is even stronger in the high-mass case \citep{stauber2004}. 
Since a collapsing protostellar envelope depends on gas-phase molecules and atoms to dissipate gravitational energy, knowledge of the chemical composition and excitation conditions of atomic and molecular gas is key in the understanding of the high-mass star formation process. 

\object{AFGL2591} is a high-mass star-forming region with a centrally heated molecular envelope. 
It is located at galactic coordinates $\ell, b = 78\fdg9, 0\fdg71$ in the Cygnus-X region, which has an average distance of 1.7\,kpc \citep{motte2007}. While the distance to AFGL2591 itself has been uncertain in the past \citep{vandertak1999,schneider2006}, in this paper we adopt a distance of 3.3\,kpc as derived from trigonometric parallax measurements of H$_2$O masers by \citet{rygl2012}. With respect to the distance of 1\,kpc often used in previous work \citep[e.g.,][]{hasegawa1995,vandertak2000jul,vandertak2005,vanderwiel2011}, the new distance changes the estimates for the mass of the dominant protostar, \mbox{VLA 3}, to $\sim$40\,$M_\odot$ and the bolometric luminosity to \pow{2}{5}\,$L_\odot$ \citep{sanna2012,jimenez-serra2012}. All linear scales mentioned in the remainder of this paper are based on a distance of 3.3\,kpc to AFGL2591, as is the numerical model description we use in Sect.~\ref{sec:modelSLEDs}. 
AFGL2591 exhibits a powerful bipolar molecular outflow extending to $>$1\arcmin\ from the central source \citep{lada1984,hasegawa1995}. At a 3.3\,kpc distance, this corresponds to a linear scale $\sim$1\,pc, which is not untypical for massive protostellar outflows \citep{beuther2002b}. A massive disk has been proposed to exist around source \mbox{VLA 3}, but with an angular extent of $<$1\arcsec\ \citep{vandertak2006,wang2012} the disk will not be resolved in the current study, which traces $\gtrsim$10\arcsec\ scales. 
Although interferometric observations have revealed multiple sources inside the envelope at spacings of $\lesssim$10\,000\,AU \citep{vandertak1999,trinidad2003,benz2007,dewit2009,sanna2012}, AFGL2591 is isolated at scales of a few parsecs. Such isolation is exceptional for massive star-forming envelopes, which are usually found in clustered environments. The absence of complex dynamical and radiative interaction with neighboring objects has made AFGL2591 a popular high-mass protostellar object to study.

Most previous modeling efforts for the large scale envelope of AFGL2591 have assumed a spherical morphology with a power law density profile \citep{vandertak1999,vandertak2000jul,doty2002,stauber2005,benz2007,dewit2009}. In these studies, gas temperatures range from $\sim$20\,K in the outskirts of the envelope ($\sim$$10^5$\,AU from the center) to several hundred K at distances as close as a few hundred AU from the protostar. Likewise, \citet{vanderwiel2011} employ spherically symmetric model geometries, but they also turn to flattened axially symmetric geometries to explain molecular emission maps at scales of $\sim$$10^4$--$10^5$\,AU. Moreover, after suggestions for the presence of cavities in the envelope carved out by outflow motions \citep{vandertak1999}, and observational evidence in the near-infrared by \citet{preibisch2003}, detailed radiative transfer models including outflow cavities are presented by \citet{bruderer2009b,bruderer2009a,bruderer2010a}. They include a chemical balance to allow for photodissociation of molecules along the interface between the envelope and the cavity, and show that a considerable fraction of the gas is at temperatures exceeding 1000\,K. This adjusted temperature balance and the possible presence of shocked regions obviously affect the excitation conditions of the molecular gas. 

In the envelopes of massive star-forming regions, much of the gas is well below 50\,K, but temperatures also exceed 100\,K in a significant portion of the gas. In this work, we study this warm gas in the large scale envelope of AFGL2591. To study these high gas temperatures with ground-based observations, one relies on low abundance species with relatively complex spectra such as \methanol, \SOtwo\ and \HHCO. Because \water\ is only abundant at temperatures above $\sim$100\,K \citep{ceccarelli1996,doty1997}, the most important probes of star-forming gas are linear rotor molecules. These species, however, are only observable from the ground in transitions that trace either very hot ($\sim$$10^3$\,K for near-infrared \HH\ lines) or cold gas ($\lesssim$50\,K for low $J$ transitions of asymmetric linear rotors in the (sub)millimeter). Space-based observations are therefore necessary to obtain a comprehensive picture of the warm gas at a few hundred~K. 
We exploit the far-infrared frequency range accessible to the {\it Herschel} Space Observatory \citep{pilbratt2010} and its Heterodyne Instrument for the Far-Infrared \citep[HIFI,][]{degraauw2010} to observe CO, \HCOplus, CS, HCN and HNC in high rotational transitions with upper level energies ranging from 80 to $\sim$700\,K. 

This paper is structured as follows. Section~\ref{sec:2591highexcobs} describes the observations and processing of spectra obtained with {\it Herschel}/HIFI from space and of complementary ground-based millimeter-wave observations. In Sect.~\ref{sec:obsresults} we present line detections of CO and isotopologues, \HCOplus, CS, HCN and HNC, and we discuss trends in line position, shape and intensity. The observed line strengths of the envelope component are compared with radiative transfer models in Sect.~\ref{sec:modelSLEDs}. Section~\ref{sec:componentcolumn} presents a derivation of physical conditions in kinematically distinct components. Finally, we discuss the results in Sect.~\ref{sec:discussion} and summarize our conclusions in Sect.~\ref{sec:summaryoutlook}.

\section{Observations}
\label{sec:2591highexcobs}

\subsection{490--1850 GHz {\it Herschel}/HIFI spectroscopy}
\label{sec:hifiobs}

The \textit{Herschel} Key Program ``Chemical \textit{HErschel} Surveys of Star-forming regions"  \citep[CHESS,][]{ceccarelli2010} uses the HIFI instrument to perform spectral surveys of various Galactic star-forming regions. One of the targets is AFGL2591, for which an unbiased spectral survey was conducted in the 490--1240\,GHz  range (bands 1a--5a, totaling 18.4 hours of observing time), complemented by selected frequency settings between 1260 and 1850 GHz (bands 5b--7b). Of the total of eight targeted frequency settings, this work only uses the five that cover $^{12}$CO transitions (2.1 hours of observing time).

\begin{table*}
	\caption{Overview of {\it Herschel}/HIFI observations of AFGL2591 used in this work.}
	\label{t:obsoverview}
\begin{tabular}{c l l r@{--} l c c r@{}l r @{--} l r}
\hline\hline 						
Band & observing date & {\it Herschel} obsid 	& \multicolumn{2}{c}{Freq.~range\tablefootmark{a}} 	& $t_\mathrm{obs}$\tablefootmark{b}	& rms noise\tablefootmark{c}  & \multicolumn{2}{c}{$T_\mathrm{sys}$\tablefootmark{d}}  & \multicolumn{2}{c}{HPBW\tablefootmark{e}} & \etamb\tablefootmark{f} \\
 & (yyyy-mm-dd) & 		& \multicolumn{2}{c}{(GHz)}	& (s)		& (mK)	     & \multicolumn{2}{c}{(K)}                             & \multicolumn{2}{c}{(\arcsec)} & (\%) \\
\hline \noalign{\smallskip}
1a	& 2010-04-11 & 1342194483 & 	483	& 558	& 4592	& 30	& $74$ & $^{+27}_{-14}$		& 44	& 38	& 75.5 \bigbr
1b	& 2010-04-12  & 1342194528 & 	555	& 636	& 4643	& 29	& $87$ & $^{+13}_{-15}$		& 38	& 33	& 75.3 \bigbr
2a	& 2010-10-28  & 1342207599 & 	631	& 722	& 9833	& 26	& $142$ & $^{+39}_{-24}$	& 34	& 29	& 75.1 \bigbr
2b	& 2010-04-12  & 1342194573 &	717	& 800	& 6407	& 67	& $188$ & $^{+56}_{-21}$	& 30	& 27	& 74.9 \bigbr
3a	& 2010-10-28  & 1342207622 &	800	& 859 	& 4893	& 39	& $200$ & $^{+88}_{-53}$	& 27 & 25	& 74.7  \bigbr
3b	& 2010-04-13  & 1342194699  & 	858	& 960	& 8578	& 67	& $205$ & $^{+148}_{-50}$	& 25	& 22	& 74.4 \bigbr
4a	& 2010-05-16  & 1342196594  &	950	& 1060	& 9137	& 157 & $356$ & $^{+294}_{-90}$ 	& 22	& 20	& 74.1 \bigbr
4b	& 2010-05-11  &  1342196428  &	1051	& 1120	& 6300	& 144 & $339$ & $^{+270}_{-63}$ 	& 20 & 19	& 73.8 \bigbr
5a	& 2010-05-12  & 1342196510  &	1110	& 1240	& 11931 	& 147 & $913$ & $^{+343}_{-215}$	& 19	& 17	& 63.7 
\bigbr
5b	& 2011-05-11  & 1342220520  & \multicolumn{2}{c}{$\sim$1267} & 1380 & 149 & 1246 & $^{\pm 126}$  & \multicolumn{2}{c}{17} & 63.4     \\ 
6a    & 2011-04-21  &  1342219224 & \multicolumn{2}{c}{$\sim$1497} & 1440 & 117 & 1507 & $^{\pm 143}$  & \multicolumn{2}{c}{14} & 71.8  \\ 
6b    & 2010-12-02  & 1342210726 & \multicolumn{2}{c}{$\sim$1612} & 1392  & 106 & 1366 & $^{\pm 104}$  & \multicolumn{2}{c}{13} & 71.1   \\ 
7a    & 2010-12-04  & 1342210803 &  \multicolumn{2}{c}{$\sim$1727} & 1575 & 92 & 1300 & $^{\pm 43}$    & \multicolumn{2}{c}{12} & 70.5  \\  
7b    & 2011-05-11  & 1342220471 & \multicolumn{2}{c}{$\sim$1841} & 1711 & 92 & 1201 & $^{\pm 5}$      & \multicolumn{2}{c}{12} & 69.7  \\ 

\hline
\end{tabular} \\
\tablefoottext{a}{Observations from bands 1a to 5a are spectral scans, those in bands 5b to 7b are single frequency settings spanning $\sim$4\,GHz.} \\
\tablefoottext{b}{Observing time, including overheads.} \\
\tablefoottext{c}{Typical rms noise level in 0.5\,MHz channels in \Tmb\ units. } \\
\tablefoottext{d}{Double sideband system temperature. For spectral scans (1a--5a): median of the individual datasets, with superscripts and subscripts denoting the deviations to the 90$^\mathrm{th}$ and 10$^\mathrm{th}$ percentiles, respectively. For pointed frequency settings (5b-7b): average of the values for the H- and V-datasets, with superscripts denoting the deviation to lowest and highest value.} \\
\tablefoottext{e}{Half-power beam width at the relevant frequency, calculated following \citet{roelfsema2012}. The two values for each spectral scan indicate the HPBW at the lower and upper edge of the band.} \\
\tablefoottext{f}{Typical value of the main beam efficiency in the center of the band.}
\end{table*}

The observed frequency bands are listed in Table~\ref{t:obsoverview}, along with observing dates, typical values for the noise level, system temperature, and half-power beam width (HPBW). Observations were carried out using the dual beam switch mode with off positions 3\arcmin\ east and west of the target position: RA\,=\,20$^\mathrm{h}$29$^\mathrm{m}$24\fs9, $\delta$\,=\,40\degr11\arcmin21\arcsec\ (J2000). The spectral scans (bands 1a--5a) were conducted with the wideband spectrometer (WBS) using a redundancy factor of 4. Per band, per polarization (H or V), per sideband (lower or upper), the scans contain 90--151 datasets, each composed of four 1\,GHz `subband' spectra. The pointed frequency settings (bands 5b--7b) were also executed in the dual beam switch mode, in this case with the fast chop and stability optimization options selected. The spectral resolution of the WBS is 1.1\,MHz, corresponding to 0.66\,\kms\ at 500\,GHz and 0.18\,\kms\ at 1850\,GHz. 

The raw data of the spectral scans were processed with the standard pipeline included in HIPE \citep{ott2010} version 8.1. 
Further processing of the individual datasets in the spectral scans was done in HIPE, using scripts written by the CHESS team \citep{kama2013}: (1) data inspection and flagging of spectral regions with spurious features, (2) subtraction of sinusoid standing waves, (3) subtraction of a polynomial baseline, generally of order $\le 3$, fitted to line-free regions.  
Strong spectral lines are known to create minor ghost lines in the sideband deconvolution process (see below). Therefore, the above steps were repeated by flagging not just the spurious features, but also the strong spectral lines, resulting in a cleaner secondary spectrum with all strong lines masked. The threshold for marking lines as strong varied from \Tastar\ $=$1\,K in band 1 to 8\,K in band 5a, depending on the amount of lines and the noise level in a particular band. 

The spectral scans of the CHESS program were executed such that each sky frequency was covered by several settings of the local oscillator, which allows for the mathematical disentanglement of the upper and lower sideband signals that result from double sideband heterodyne observations \citep{comito2002}. This `sideband deconvolution' was performed with sideband gain ratios fixed to unity. Spectra from the HIFI spectral scans (490--1240\,GHz) presented in this paper are all single sideband, `unfolded' spectra. 

The correspondence of deconvolved products for the two polarization backends, H and V, is checked and found to agree to within 3\% in peak intensity for various lines, and line peak positions and line shapes generally match to within the noise level (cf.~Table~\ref{t:obsoverview}). The signal from the two polarizations is averaged by feeding the datasets from the two backends to the sideband deconvolution algorithm simultaneously. One exception is band 4b, where we used only the H polarization, since V is contaminated by numerous spurious features and the V-only deconvolved spectrum exhibits noise levels of up to four times higher than in the H polarization. 


The WBS spectra from the pointed frequency settings of CO transitions above 1240\,GHz (bands 5b, 6a, 6b, 7a and 7b) are processed with HIPE version 7. Subsequent removal of polynomial baselines of order $\leq$4 and sine-shaped standing waves is applied in the CLASS package\footnote{CLASS is part of the GILDAS software developed at IRAM: \texttt{http://www.iram.fr/IRAMFR/GILDAS}.}. With a single frequency setting, sideband deconvolution is impossible, so the high-frequency CO lines are double sideband spectra. Like in most spectral scans described above, the average of the two polarizations is used, with the exception of CO 11--10 in band 5b. For this line, the spectral line shape shows discrepancies between H and V. Since the baseline rms noise level is $\sim$20\% higher in V, we only use the H polarization in this case. 

We use the main beam efficiency \etamb, as determined from a beam characterization experiment based on Mars observations \citep{roelfsema2012}, to convert the \Taprime\ intensity scale to main beam intensity: \Tmb~=~$(\eta_\mathrm{fwd}/\eta_{mb})$\Tastar\ with the forward efficiency $\eta_\mathrm{fwd}=0.96$ for HIFI. Although \etamb\ is in principle frequency-dependent, it varies by $<1\%$ across an individual band. A more significant variation in the value of \etamb\ is the 14\% drop between band 4b and 5a (Table~\ref{t:obsoverview}). 

The scaling to main beam intensity is strictly only applicable to emission regions that are not larger than the main beam of the telescope. For the lines studied here, this criterion is largely obeyed, so we apply the \etamb\ factor to facilitate a direct comparison to modeled intensities. 

Conservative estimates put the formal uncertainty for the intensity calibration of HIFI at 9\% in bands 1 and 2, at 10\% in bands 3 and 4, and $\sim$14\% in bands 5--7 \citep{roelfsema2012}. However, the uncertainty budget is often dominated by standing waves and the uncertainty in the sideband ratio, both of which are mitigated considerably by our systematic cleaning (all bands) and sideband deconvolution of the spectral scans (bands 1a--5a). Based on consistency checks of HIFI line measurements over {\it Herschel}'s operational lifetime, the observed depth of saturated absorption lines, comparison to ground-based observations and cross-comparison with {\it Herschel}/SPIRE, realistic values for HIFI's absolute calibration uncertainty should be between 5 and 10\%. We adopt a uniform uncertainty of 10\% for the line intensities measured from our HIFI spectra.

\subsection{Ground-based (sub)-millimeter spectroscopy}

Apart from the {\it Herschel} spectra, this paper uses ground-based observations with the IRAM 30m Telescope\footnote{IRAM is supported by INSU/CNRS (France), MPG (Germany) and IGN (Spain).} and the James Clerk Maxwell Telescope\footnote{The James Clerk Maxwell Telescope is operated by the Joint Astronomy Centre on behalf of the Science and Technology Facilities Council of the United Kingdom, the Netherlands Organisation for Scientific Research, and the National Research Council of Canada.} (JCMT). 

IRAM 30m observations were performed on July 21, 2010. The observations were conducted with the EMIR0 and EMIR2 receivers in parallel (two linear polarizations each) with the VESPA autocorrelator backend in the 2.7\,mm ($\sim$110\,GHz) and 1.3\,mm ($\sim$230\,GHz) windows\footnote{See \url{http://www.iram-institute.org/} for instrument specifications.}. The \CeiO\ and \thCO\ $J$=1--0  lines were observed simultaneously by splitting the backend in two subwindows centered on the lines with a frequency sampling of 78 kHz ($\sim$0.2\,\kms). The \CeiO~2--1 line was observed with the same frequency sampling ($\sim$0.1\,\kms). We used frequency switching mode and an integration time of 1 minute on the sky (2 minutes after averaging the two polarizations). The weather was good with precipitable water vapor of 3\,mm, and system temperature was 100\,K at 2.7\,mm and 200\,K at 1.3\,mm, in ${T_A}^*$ scale. A fourth order polynomial baseline has been subtracted. The measured noise in the spectra, in \Tmb\ units, is 46\,mK for the 2.7\,mm spectra and 116\,mK for the 1.3\,mm spectrum after averaging and folding. We use HPBW of 22\arcsec\ and 11\arcsec, and main beam efficiencies (\etamb) of 78\% and 61\% for 2.7\,mm and 1.3\,mm, respectively. For the IRAM/EMIR spectra, we take an absolute intensity calibration uncertainty of 10\%.

Spectra of rotational transitions of CO species, HCN, HNC, \HCOplus, and CS that fall between 330 and 373\,GHz are extracted from JCMT Spectral Legacy Survey data \citep{plume2007,vanderwiel2011}. The beam size of the JCMT is \mbox{14--15\arcsec}\ at these frequencies, and the spectral resolution used in the survey is 1\,MHz, corresponding to $\sim$0.8\,\kms. The absolute intensity calibration for the JCMT measurements is uncertain to 15\%. A detailed description of the data acquisition and processing is given by \citet{vanderwiel2011}.


\begin{figure}
	 \resizebox{\hsize}{!}{\includegraphics{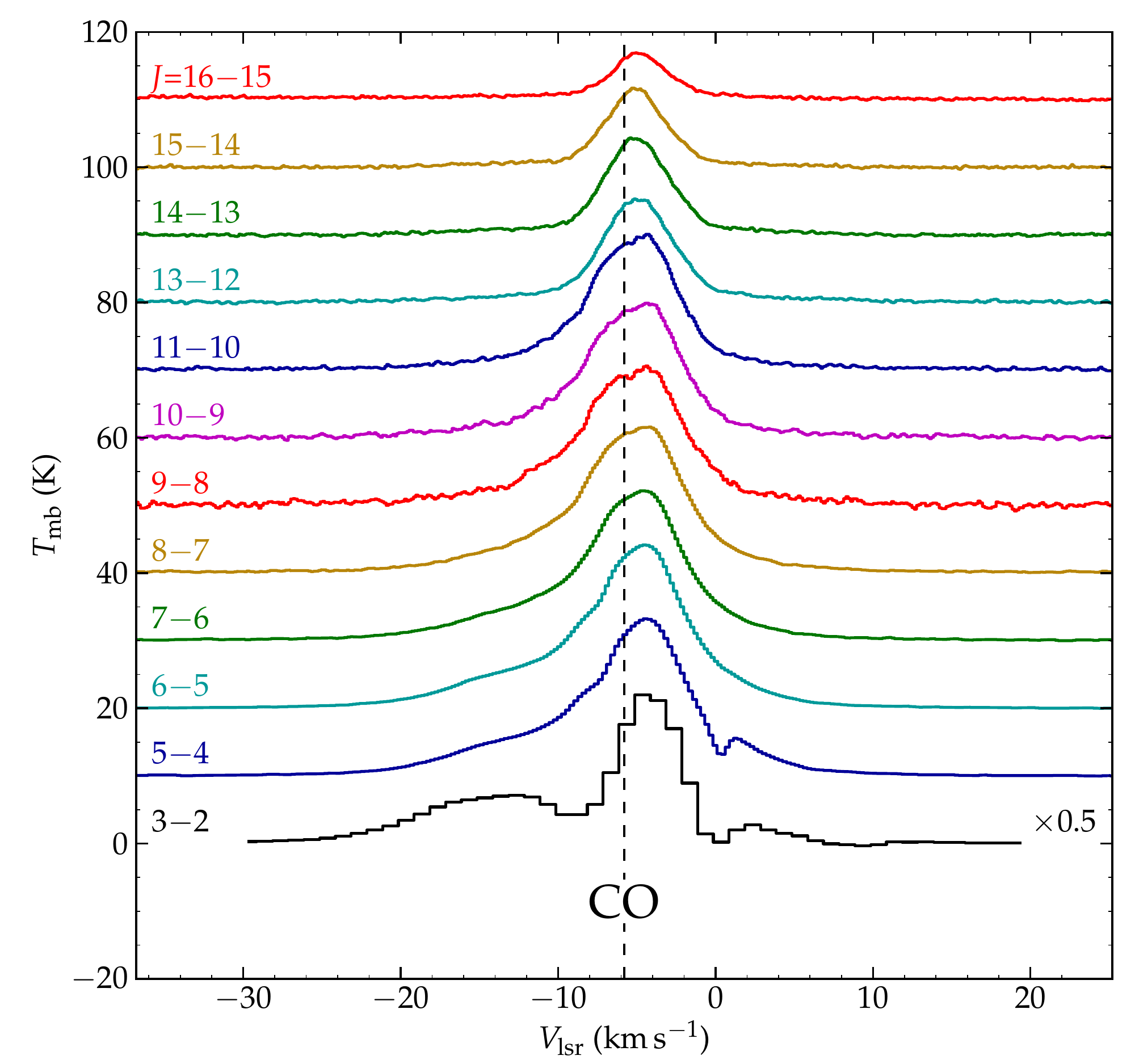}}
  	 \resizebox{\hsize}{!}{\includegraphics{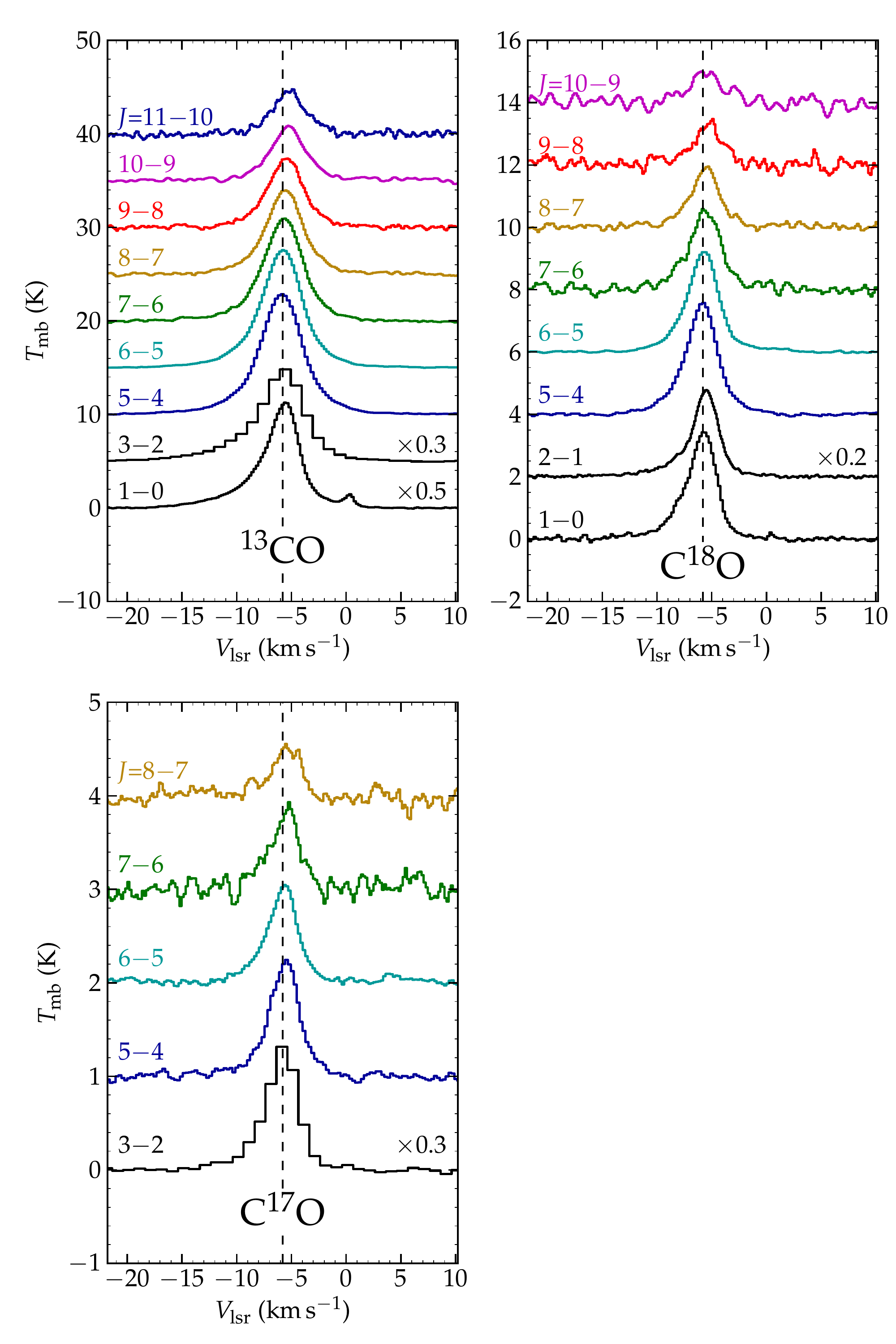}}
	\caption{Spectral line profiles of CO, \thCO, \CeiO, and \CstO. The \mbox{1--0} and \mbox{2--1} spectra are from the IRAM 30m telescope, the 3--2 lines are from JCMT/HARP-B. All higher-$J$ lines are from {\it Herschel}/HIFI. For clarity, consecutive spectra, each labeled by their upper and lower $J$-level, are offset in the vertical direction. Corresponding rest frequencies and energy levels are listed in Table~\ref{t:obslines}. The systemic velocity ($-5.8$\,\kms) of AFGL2591 is indicated by the dashed vertical line.   }
	\label{fig:CO_lineprofiles}
\end{figure}

\begin{figure}
	\resizebox{\hsize}{!}{\includegraphics{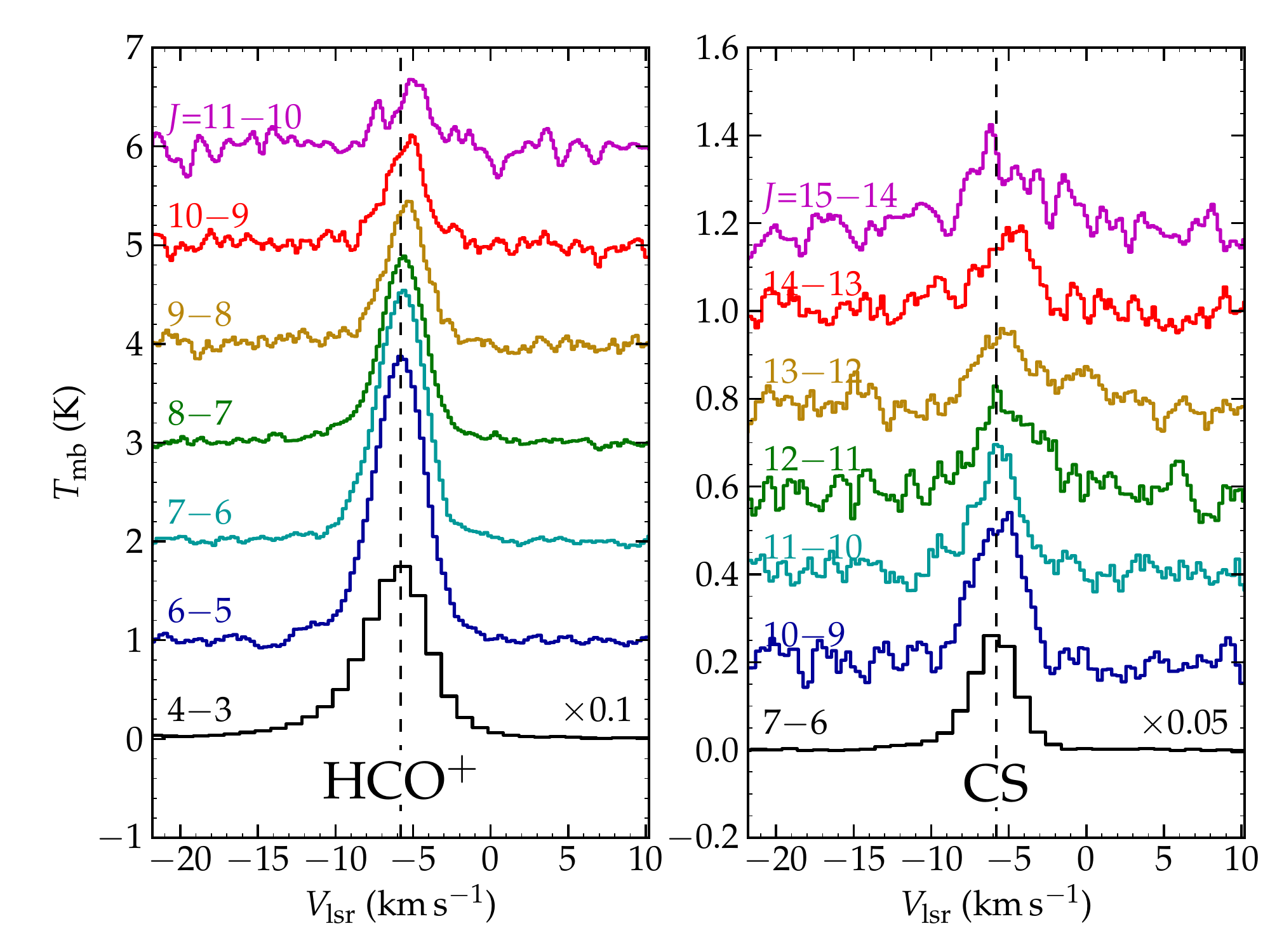}}
	\resizebox{\hsize}{!}{\includegraphics{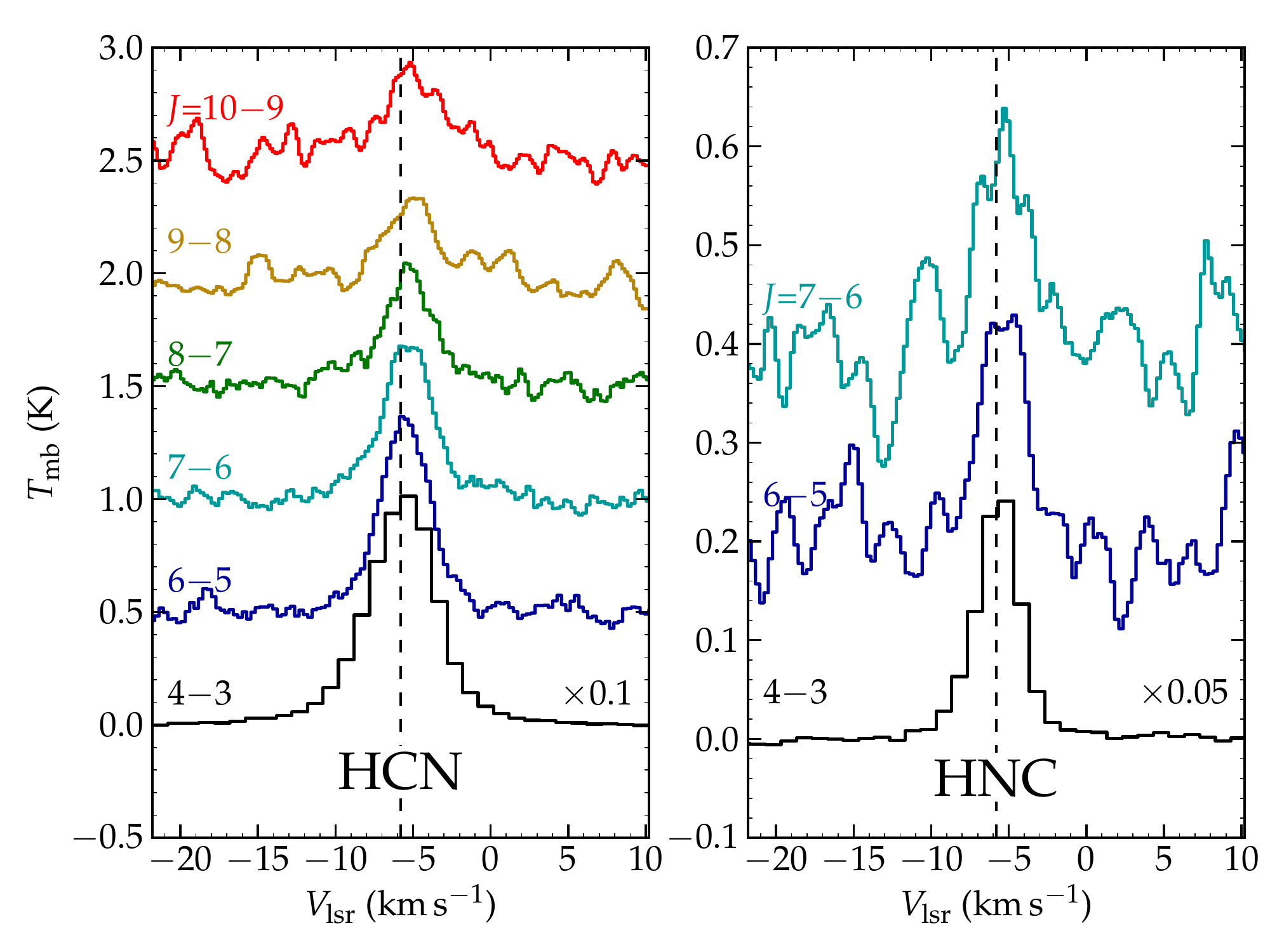}}
	\caption{As Fig.~\ref{fig:CO_lineprofiles}, but for \HCOplus, CS, HCN, and HNC. All lines are measured with {\it Herschel}/HIFI, except \HCOplus, HCN, and HNC 4--3, as well as CS 7--6, which are measured with JCMT. }
	\label{fig:nonCO_lineprofiles}
\end{figure}

\section{Observational results}
\label{sec:obsresults}

\subsection{Detected emission lines and line shapes}
\label{sec:obsdetected}

The spectral scans and targeted frequency settings described in Sect.~\ref{sec:2591highexcobs} cover high-$J$ transitions of the linear molecules \HCOplus, CS, HCN, HNC, CO, \thCO, \CeiO, and \CstO, as listed in Table~\ref{t:obslines}. Other detected spectral lines will be presented in a companion paper by \citet[][hereafter `\citetalias{kazmierczak_inprep}']{kazmierczak_inprep}. 

For all but one of the species mentioned above, frequencies of rotational transitions are extracted from the Cologne Database for Molecular Spectroscopy \citep[CDMS,][]{muller2005}. For \CstO\ we use the JPL frequencies \citep{pickett1998}\footnote{The JPL database lists an average frequency for each rotational transition of \CstO, whereas CDMS lists all 15 hyperfine transitions separately, which -- at a maximum separation of $\sim$2\,MHz -- are blended in all \CstO\ lines observed in this work.}. We search the processed HIFI spectra for line signal near these frequencies. 
For weak lines (peak intensity $<5\sigma_\mathrm{rms}$) the spectrum is smoothed in steps of a factor 2, until channels become so wide that they become comparable to the line width. If the $S/N$ is still below 5 at this point, a conservative upper limit to the integrated line flux is then set at $5 \sigma_\mathrm{rms}~\times$~(3\,\kms), with the last factor motivated by the typical line width expected from firmly detected lines at lower $J$. 

Spectral line profiles of transitions of  CO, \thCO, \CeiO, and \CstO\ are presented in Fig.~\ref{fig:CO_lineprofiles}; those of \HCOplus, CS, HCN, and HNC are shown in Fig.~\ref{fig:nonCO_lineprofiles} (see also Table~\ref{t:obslines}). If a line needs smoothing of the spectrum before its peak intensity is above 5$\sigma_\mathrm{rms}$, Fig.~\ref{fig:nonCO_lineprofiles} gives the smoothed spectrum. 

All lines with sufficient signal-to-noise show an asymmetric profile with line wings that are stronger on the blueshifted than on the redshifted side. 
The blueshifted line wing (\vlsr~$\lesssim -10$\,\kms) is visible in emission in the lowest HIFI transitions of CO, and to a lesser degree of \HCOplus. We attribute the asymmetry on the blue side of the line profile to the approaching outflow (see Sect.~\ref{sec:obsSLEDs}). In both CO and \HCOplus, it becomes less pronounced at higher $J$-levels, but it is nevertheless present up to the highest CO transition observed ($J$=16--15). A separation between envelope and outflow emission is also seen in CO~2--1 spectra by \citet{mitchell1992}, similar to our CO~\mbox{3--2} profile. In addition, this line wing is known from infrared absorption studies of $^{12}$CO and \thCO. In $^{12}$CO, the outflow is seen out to 200\,\kms\ blueshifted \citep{mitchell1989,vandertak1999}. 

Contrary to most bright lines, which peak at the systemic velocity and have a blueshifted line wing, the peaks of all $^{12}$C$^{16}$O lines (Fig.~\ref{fig:CO_lineprofiles}) are displaced toward redshifted velocities. This is attributed to the optical depth and the possible presence of absorption components affecting the line shape. Both these effects scale with the overall abundance, which is far higher for $^{12}$C$^{16}$O than for any other species studied here. 
In addition, the 12--11 and 14--13 lines of CS (Fig.~\ref{fig:nonCO_lineprofiles}) appear asymmetric with excess emission on the redshifted side at the $\sim$2-$\sigma$ level.

The 11--10 line of \HCOplus\ appears to be split in two separate components. The brightest one follows the trend of increasingly redder peak velocity with higher energies, while a secondary, narrow peak emerges at $-8$\,\kms\ (5\,$\sigma$ detection). The same feature, although fainter than for $J$=11--10, is visible in the 10--9 and possibly also the 9--8 lines of \HCOplus\ when a single component profile based on lower lines is subtracted. In the other species studied here, this additional narrow component is found only in \CstO\ 8--7 at the 2.5-$\sigma$ level, while it might be hidden in the line wing of \CstO\ 7--6. Based on the firm detection in \HCOplus, we conclude that the feature must originate from dense and very warm gas, given that it is brighter at $J$=11--10 than at lower transitions.

\begin{sidewaystable*}
	\vspace{0.0\textwidth}  
	\caption{Molecular line measurements.}
	\label{t:obslines}
	\begin{tabular}{@{} l @{\ } l r r r @{ $\pm$ } l r@{ $\pm$ } l r@{ $\pm$ } l @{\extracolsep{1em}} r @{\extracolsep{0em}} @{ $\pm$ } l r@{ $\pm$ } l r@{ $\pm$ } l  @{}}   
  	\hline
	\hline
	 & & & & \multicolumn{6}{c}{Envelope component\tablefootmark{b}} & \multicolumn{6}{c}{Outflow component\tablefootmark{b}} \\
	 \cline{5-10} 
	 \cline{11-16}
Molecule & $J_\mathrm{up}$--$J_\mathrm{low}$  & Frequency\tablefootmark{a} &  $E_\mathrm{up}/k$ &  \multicolumn{2}{c}{\intintens\tablefootmark{c}}	& \multicolumn{2}{c}{ $V_\mathrm{centroid}$\tablefootmark{d} } & \multicolumn{2}{c}{FWHM\tablefootmark{e} }  & \multicolumn{2}{c}{\intintens\tablefootmark{c}}	& \multicolumn{2}{c}{ $V_\mathrm{centroid}$\tablefootmark{d} } & \multicolumn{2}{c}{FWHM\tablefootmark{e} }\\
	& & (MHz)	& (K)	&  \multicolumn{2}{c}{(K\,\kms)} &  \multicolumn{2}{c}{(\kms)} &  \multicolumn{2}{c}{(\kms)} &  \multicolumn{2}{c}{(K\,\kms)} &  \multicolumn{2}{c}{(\kms)} &  \multicolumn{2}{c}{(\kms)} \\
	 \hline
  CO  & 3--2\tablefoottext{f} & 345795.99 & 33.2 &  195.3 & 0.2 & $-4.16$ & 0.001 & 4.133 & 0.003 & 170.8 & 0.4 & $-14.031$ & 0.008 & 11.40 & 0.02  \\
   CO & 5--4 &  576267.93 & 83.0 & 76.6 & 0.1 & $-4.42$ &  0.02 &  4.98 &  0.05 
		& 155.7 & 0.2 & $-6.70$ &  0.06 & 15.85 &  0.12 \\
   CO & 6--5 &  691473.08 & 116.2 & 92.1 & 0.1 & $-4.61$ &  0.01 &  5.39 &  0.04 
	      	& 145.5 & 0.1 & $-7.23$ &  0.05 & 15.70 &  0.10 \\
   CO & 7--6 &  806651.81 & 154.9 & 90.8 & 0.1 & $-4.78$ &  0.01 &  5.52 &  0.03 
	       	& 119.5 & 0.1 & $-7.36$ &  0.04 & 15.09 &  0.09 \\
   CO & 8--7 &  921799.70 & 199.1 & 100.1 & 0.2 & $-4.90$ &  0.02 &  5.96 &  0.06 
		& 105.9 & 0.2 & $-7.21$ &  0.09 & 15.17 &  0.18 \\
   CO & 9--8 &  1036912.39 & 248.9 & 109.9 & 0.3 & $-5.02$ &  0.03 &  6.43 &  0.09 
		& 77.8 & 0.5 & $-7.22$ &  0.18 & 15.96 &  0.42 \\
   CO & 10--9 &  1151985.45 & 304.2 & 102.9 & 0.2 & $-4.97$ &  0.02 &  6.01 &  0.07 
		& 66.1 & 0.4 & $-7.29$ &  0.17 & 14.64 &  0.35 \\
   CO & 11--10 &  1267014.49 & 365.0 & 100.8 & 0.2 & $-5.05$ &  0.02 &  5.71 &  0.06 
		& 58.1 & 0.4 & $-7.21$ &  0.17 & 14.87 &  0.40 \\
   CO & 13--12 &  1496922.91 & 503.1 & 69.5 & 0.1 & $-4.94$ &  0.01 &  4.82 &  0.03 
     & 34.7 & 0.8 & $-4.95$ &  0.30 & 18.24 &  0.83 \\
   CO & 14--13 &  1611793.52 & 580.5 & 62.2 & 0.1 & $-5.00$ &  0.01 &  4.54 &  0.02 
     & 24.7 & 0.4 & $-5.98$ &  0.11 & 17.47 &  0.36 \\
   CO & 15--14 &  1726602.51 & 663.4 & 46.6 & 0.1 & $-5.08$ &  0.01 &  4.22 &  0.02 
     & 18.0 & 0.4 & $-6.07$ &  0.13 & 16.61 &  0.41 \\      
   CO & 16--15 &  1841345.51 & 751.7 & 23.1 & 0.1 & $-4.82$ &  0.01 &  3.80 &  0.03 
     & 12.1 & 0.4 & $-4.96$ &  0.11 & 13.34 &  0.40 \\
\hline 
\thCO & 1--0\tablefoottext{g} &  110201.35 & 5.3 & 45.7 & 0.3 & $-5.57$ &  0.02 &  2.71 &  0.06  & 57.3 & 0.3 & $-7.09$ &  0.09 &  7.23 &  0.15 \\ 
\thCO & 3--2\tablefoottext{f} & 330587.97 & 31.7 &  92.3 & 0.6 & $-5.578$ & 0.004 & 3.68 & 0.02 & 97.3 & 1.1 & $-7.33$ & 0.03 & 9.13 & 0.04  \\
\thCO & 5--4 &  550926.29 & 79.3 & 38.7 & 0.1 & $-5.85$ &  0.01 &  3.78 &  0.03 
		      & 31.6 & 0.2 & $-6.34$ &  0.04 &  9.06 &  0.14 \\
\thCO & 6--5 &  661067.28 & 111.1 & 34.6 & 0.1 & $-5.76$ &  0.01 &  3.58 &  0.03 
		      & 31.0 & 0.2 & $-6.41$ &  0.03 &  8.21 &  0.11 \\
\thCO & 7--6 &  771184.12 & 148.1 & 32.1 & 0.2 & $-5.66$ &  0.01 &  3.63 &  0.04 
		      & 22.4 & 0.2 & $-6.34$ &  0.06 &  8.06 &  0.19 \\
\thCO & 8--7 &  881272.81 & 190.4 & 21.0 & 0.2 & $-5.55$ &  0.01 &  3.18 &  0.05 
			& 20.8 & 0.2 & $-5.92$ &  0.04 &  7.03 &  0.17 \\
\thCO & 9--8 &  991329.31 & 237.9 & 14.1 & 0.3 & $-5.43$ &  0.02 &  2.94 &  0.10 
			& 18.5 & 0.4 & $-5.77$ &  0.05 &  6.03 &  0.21 \\
\thCO & 10--9 &  1101349.60 & 290.8 & 6.2 & 0.4 & $-5.33$ &  0.04 &  2.32 &  0.18 
		      & 18.1 & 0.4 & $-5.41$ &  0.04 &  5.19 &  0.21 \\
\thCO & 11--10 &  1211329.66 & 348.9 & 17.4 & 0.1 & $-5.22$ &  0.03 &  3.81 &  0.07 
	& \multicolumn{2}{c}{-} & \multicolumn{2}{c}{-} & \multicolumn{2}{c}{-}  \\
\hline
\CeiO & 1--0\tablefoottext{g} &  109782.17 & 5.3  & 5.3 & 0.1 & $-5.57$ &  0.02 &  2.24 &  0.08 
 & 7.0 & 0.2 & $-6.68$ &  0.10 &  4.64 &  0.12 \\
\CeiO & 2--1\tablefoottext{g} &  219560.35 & 15.8  & 25.9 & 0.1 & $-5.45$ &  0.01 &  2.26 &  0.03 
 & 22.1 & 0.1 & $-6.68$ &  0.06 &  6.13 &  0.10 \\
\CeiO & 5--4 &  548831.01 & 79.0 & 6.2 & 0.1 & $-5.79$ &  0.01 &  2.55 &  0.05 
			& 7.8 & 0.1 & $-6.30$ &  0.04 &  5.72 &  0.12 \\
\CeiO & 6--5 &  658553.28 & 110.6 & 6.3 & 0.1 & $-5.69$ &  0.01 &  2.63 &  0.06 
		      & 6.2 & 0.2 & $-6.23$ &  0.06 &  5.84 &  0.19 \\
\CeiO & 7--6 &  768251.59 & 147.5 & 4.6 & 0.3 & $-5.47$ &  0.05 &  2.58 &  0.19 
			& 5.3 & 0.5 & $-6.01$ &  0.17 &  5.71 &  0.51 \\
\CeiO & 8--7 &  877921.95 & 189.6 & 1.7 & 0.2 & $-5.41$ &  0.04 &  1.79 &  0.15 
		      & 4.9 & 0.2 & $-5.82$ &  0.06 &  4.32 &  0.18 \\
\CeiO & 9--8 &  987560.38 & 237.0 & 4.1 & 0.1 & $-5.30$ &  0.03 &  3.24 &  0.06 
	& \multicolumn{2}{c}{-} & \multicolumn{2}{c}{-} & \multicolumn{2}{c}{-}  \\
\CeiO & 10--9 &  1097162.88 & 289.7 & 3.8 & 0.2 & $-5.51$ &  0.08 &  4.00 &  0.19 
	& \multicolumn{2}{c}{-} & \multicolumn{2}{c}{-} & \multicolumn{2}{c}{-}  \\
\CeiO & 11--10 & 1206725.45 & 347.6 & \multicolumn{2}{c}{ $<$ 2.0 }  
	& \multicolumn{2}{c}{-}& \multicolumn{2}{c}{-}& \multicolumn{2}{c}{-}& \multicolumn{2}{c}{-}& \multicolumn{2}{c}{-}  \\
\hline
\CstO & 3--2\tablefoottext{f} & 337061.15 & 32.4 & 9.8 & 0.6 & $-5.64$ & 0.04 & 2.8 & 0.1 & 7.3 & 1.0 & $-6.9$ & 0.2 & 5.7 & 0.3  \\
\CstO & 5--4 &  561712.78 & 80.9 & 4.3 & 0.1 & $-5.63$ &  0.03 &  3.44 &  0.07 
	& \multicolumn{2}{c}{-} & \multicolumn{2}{c}{-} & \multicolumn{2}{c}{-}  \\
\CstO & 6--5 &  674009.34 & 113.2 & 3.6 & 0.1 & $-5.81$ &  0.03 &  3.51 &  0.08 
	& \multicolumn{2}{c}{-} & \multicolumn{2}{c}{-} & \multicolumn{2}{c}{-}  \\
\CstO & 7--6 &  786280.82 & 151.0 & 2.8 & 0.2 & $-5.60$ &  0.07 &  3.37 &  0.16 
	& \multicolumn{2}{c}{-} & \multicolumn{2}{c}{-} & \multicolumn{2}{c}{-}  \\
\CstO & 8--7 &  898523.02 & 194.1 & 1.7 & 0.1 & $-5.44$ &  0.06 &  3.04 &  0.15 
	& \multicolumn{2}{c}{-} & \multicolumn{2}{c}{-} & \multicolumn{2}{c}{-}  \\
\CstO & 9--8 & 1010731.78 & 242.6 & \multicolumn{2}{c}{ $<$1.3 }  
	& \multicolumn{2}{c}{-}& \multicolumn{2}{c}{-}& \multicolumn{2}{c}{-}& \multicolumn{2}{c}{-}& \multicolumn{2}{c}{-}  \\

\hline
\multicolumn{16}{c}{(continued on next page)}
\end{tabular}
\end{sidewaystable*}
\addtocounter{table}{-1}
\begin{sidewaystable*}
\caption{(continued).}
	\begin{tabular}{@{} l @{\ } l r r r @{ $\pm$ } l r@{ $\pm$ } l r@{ $\pm$ } l @{\extracolsep{1em}} r @{\extracolsep{0em}} @{ $\pm$ } l r@{ $\pm$ } l r@{ $\pm$ } l  @{}}   
  	\hline
	\hline
	 & & & & \multicolumn{6}{c}{Envelope component\tablefootmark{b}} & \multicolumn{6}{c}{Outflow component\tablefootmark{b}} \\
	 \cline{5-10} 
	 \cline{11-16}
Molecule & $J_\mathrm{up}$--$J_\mathrm{low}$  & Frequency\tablefootmark{a} &  $E_\mathrm{up}/k$ &  \multicolumn{2}{c}{\intintens\tablefootmark{c}}	& \multicolumn{2}{c}{ $V_\mathrm{centroid}$\tablefootmark{d} } & \multicolumn{2}{c}{FWHM\tablefootmark{e} }  & \multicolumn{2}{c}{\intintens\tablefootmark{c}}	& \multicolumn{2}{c}{ $V_\mathrm{centroid}$\tablefootmark{d} } & \multicolumn{2}{c}{FWHM\tablefootmark{e} }\\
	& & (MHz)	& (K)	&  \multicolumn{2}{c}{(K\,\kms)} &  \multicolumn{2}{c}{(\kms)} &  \multicolumn{2}{c}{(\kms)} &  \multicolumn{2}{c}{(K\,\kms)} &  \multicolumn{2}{c}{(\kms)} &  \multicolumn{2}{c}{(\kms)} \\
	 \hline

  HCN & 4--3\tablefoottext{f} & 354505.48  & 42.5 & 36.2 & 0.7 & $-5.51$ & 0.02 & 4.33 & 0.04 & 22.4 & 1.3 & $-6.52$ & 0.06 & 8.4 & 0.2 \\
  HCN & 6--5 &  531716.35 & 89.3 & 3.5 & 0.1 & $-5.52$ &  0.02 &  3.91 &  0.05 
  	& \multicolumn{2}{c}{-} & \multicolumn{2}{c}{-} & \multicolumn{2}{c}{-}  \\
  HCN & 7--6 &  620304.00 & 119.1 & 2.9 & 0.1 & $-5.35$ &  0.04 &  4.02 &  0.09 
  	& \multicolumn{2}{c}{-} & \multicolumn{2}{c}{-} & \multicolumn{2}{c}{-}  \\
  HCN & 8--7 &  708877.01 & 153.1 & 2.1 & 0.1 & $-5.30$ &  0.06 &  4.16 &  0.14 
  	& \multicolumn{2}{c}{-} & \multicolumn{2}{c}{-} & \multicolumn{2}{c}{-}  \\
  HCN & 9--8 &  797433.26 & 191.4 & 1.2 & 0.2 & $-5.14$ &  0.07 &  3.46 &  0.17 
  	& \multicolumn{2}{c}{-} & \multicolumn{2}{c}{-} & \multicolumn{2}{c}{-}  \\
  HCN & 10--9 &  885970.69 & 233.9 & 2.1 & 0.2 & $-4.98$ &  0.09 &  5.47 &  0.20 
	& \multicolumn{2}{c}{-} & \multicolumn{2}{c}{-} & \multicolumn{2}{c}{-}  \\
  HCN & 11-10 & 974487.20 & 280.7 & \multicolumn{2}{c}{ $<$1.9 }  
	& \multicolumn{2}{c}{-}& \multicolumn{2}{c}{-}& \multicolumn{2}{c}{-}& \multicolumn{2}{c}{-}& \multicolumn{2}{c}{-}  \\
\hline
  HNC & 4--3\tablefoottext{f} & 362630.30 & 43.5 & 8.2 & 0.7 & $-5.51$ & 0.03 & 2.5 & 0.1 & 10 & 2 & $-6.01$ & 0.07 & 4.8 & 0.2 \\
  HNC & 6--5 &  543897.55 & 91.4 & 0.9 & 0.1 & $-5.37$ &  0.06 &  3.36 &  0.14 
  	& \multicolumn{2}{c}{-} & \multicolumn{2}{c}{-} & \multicolumn{2}{c}{-}  \\
  HNC & 7--6 &  634510.83 & 121.8 & 0.8 & 0.2 & $-5.32$ &  0.08 &  3.55 &  0.19 
	& \multicolumn{2}{c}{-} & \multicolumn{2}{c}{-} & \multicolumn{2}{c}{-}  \\
  HNC & 8--7 &  725107.34 & 156.6 & \multicolumn{2}{c}{ $<$1.1 } 
 	& \multicolumn{2}{c}{-}& \multicolumn{2}{c}{-}& \multicolumn{2}{c}{-}& \multicolumn{2}{c}{-}& \multicolumn{2}{c}{-}  \\
\hline
\HCOplus & 4--3\tablefoottext{f} & 356734.22 & 42.8 & 61.7 & 0.5 & $-5.87$ & 0.01 & 4.18 & 0.02 & 39.8 & 0.9  & $-7.61$ & 0.04 & 9.46 & 0.08 \\
\HCOplus & 6--5 &  535061.58 & 89.9 & 8.5 & 0.2 & $-5.73$ &  0.03 &  3.48 &  0.10 
		       & 4.0 & 0.4 & $-6.65$ &  0.21 &  6.12 &  0.37 \\
\HCOplus & 7--6 &  624208.36 & 119.8 & 6.0 & 0.2 & $-5.55$ &  0.04 &  3.15 &  0.14 
		      & 4.5 & 0.3 & $-6.42$ &  0.20 &  5.15 &  0.27 \\
\HCOplus & 8--7 &  713341.23 & 154.1 & 5.4 & 0.0 & $-5.61$ &  0.01 &  3.10 &  0.04 
		      & 2.2 & 0.4 & $-6.62$ &  0.14 &  8.29 &  0.38 \\
\HCOplus & 9--8 &  802458.20 & 192.6 & 4.8 & 0.1 & $-5.47$ &  0.03 &  3.27 &  0.06 
	& \multicolumn{2}{c}{-} & \multicolumn{2}{c}{-} & \multicolumn{2}{c}{-}  \\
\HCOplus & 10--9 &  891557.29 & 235.4 & 3.4 & 0.1 & $-5.47$ &  0.04 &  3.11 &  0.09 
	& \multicolumn{2}{c}{-} & \multicolumn{2}{c}{-} & \multicolumn{2}{c}{-}  \\
\HCOplus & 11--10 &  980636.49 & 282.4 & 2.1 & 0.3 & $-5.21$ &  0.10 &  3.52 &  0.25 
	& \multicolumn{2}{c}{-} & \multicolumn{2}{c}{-} & \multicolumn{2}{c}{-}  \\
\HCOplus & 12--11 & 1069693.89 & 333.8 & \multicolumn{2}{c}{ $<$2.8 }  
 	& \multicolumn{2}{c}{-}& \multicolumn{2}{c}{-}& \multicolumn{2}{c}{-}& \multicolumn{2}{c}{-}& \multicolumn{2}{c}{-}  \\
\hline
   CS & 7--6\tablefoottext{f} & 342882.85 & 65.8 & 15.5 & 0.5 & $-5.80$ & 0.02 & 3.10 & 0.06 & 5.0 & 0.8 & $-7.3$ & 0.3 & 6.4 & 0.3 \\ 
   CS & 10--9 &  489750.92 & 129.3 & 1.3 & 0.1 & $-5.68$ &  0.06 &  3.69 &  0.14 
   	& \multicolumn{2}{c}{-} & \multicolumn{2}{c}{-} & \multicolumn{2}{c}{-}  \\
   CS & 11--10 &  538689.00 & 155.1 & 1.0 & 0.2 & $-5.71$ &  0.07 &  3.53 &  0.17 
   	& \multicolumn{2}{c}{-} & \multicolumn{2}{c}{-} & \multicolumn{2}{c}{-}  \\
   CS & 12--11 &  587616.48 & 183.3 & 0.9 & 0.3 & $-5.04$ &  0.11 &  4.52 &  0.25 
   	& \multicolumn{2}{c}{-} & \multicolumn{2}{c}{-} & \multicolumn{2}{c}{-}  \\
   CS & 13--12 &  636532.46 & 213.9 & 0.6 & 0.3 & $-5.24$ &  0.12 &  4.04 &  0.29 
   	& \multicolumn{2}{c}{-} & \multicolumn{2}{c}{-} & \multicolumn{2}{c}{-}  \\
   CS & 14--13 &  685435.92 & 246.8 & 0.7 & 0.3 & $-5.05$ &  0.12 &  3.93 &  0.28 
   	& \multicolumn{2}{c}{-} & \multicolumn{2}{c}{-} & \multicolumn{2}{c}{-}  \\
   CS & 15--14 &  734325.93 & 282.0 & 0.7 & 0.4 & $-5.66$ &  0.16 &  4.17 &  0.38 
	& \multicolumn{2}{c}{-} & \multicolumn{2}{c}{-} & \multicolumn{2}{c}{-}  \\
   CS & 16--15 & 783201.51 & 319.6 & \multicolumn{2}{c}{ $<$0.82 }  
 	& \multicolumn{2}{c}{-}& \multicolumn{2}{c}{-}& \multicolumn{2}{c}{-}& \multicolumn{2}{c}{-}& \multicolumn{2}{c}{-}  \\

\hline
	\end{tabular} 
	 \tablefoot{All lines are observed with {\it Herschel}/HIFI, unless a footnote indicates otherwise.} \\
	 \tablefoottext{a}{Rest frequency of the transition, taken from the Cologne Database for Molecular Spectroscopy \citep[CDMS,][]{muller2005}, except for \CstO, for which the listed frequencies are weighted averages of the individual hyperfine transitions taken directly from JPL \citep{pickett1998}. } \\
	 \tablefoottext{b}{If two fit components are shown for one transition, the narrowest is attributed to the envelope and the other to the outflow. If only one component is fitted, it is attributed to the envelope.} \\
	\tablefoottext{c}{Integrated line intensity measured by a Gaussian fit. Upper limits for non-detections are marked by ``$<$" signs.} \\
	\tablefoottext{d}{Centroid velocity in the LSR frame, derived from a Gaussian line profile fit. } \\
	\tablefoottext{e}{Gaussian FWHM.} \\
	\tablefoottext{f}{Observed with JCMT/HARP-B \citep{vanderwiel2011}.} \\
	\tablefoottext{g}{Observed with IRAM 30m/EMIR.}
\end{sidewaystable*}

\subsection{Observed spectral line energy distributions}
\label{sec:obsSLEDs}

\begin{figure}
	\resizebox{\hsize}{!}{\includegraphics{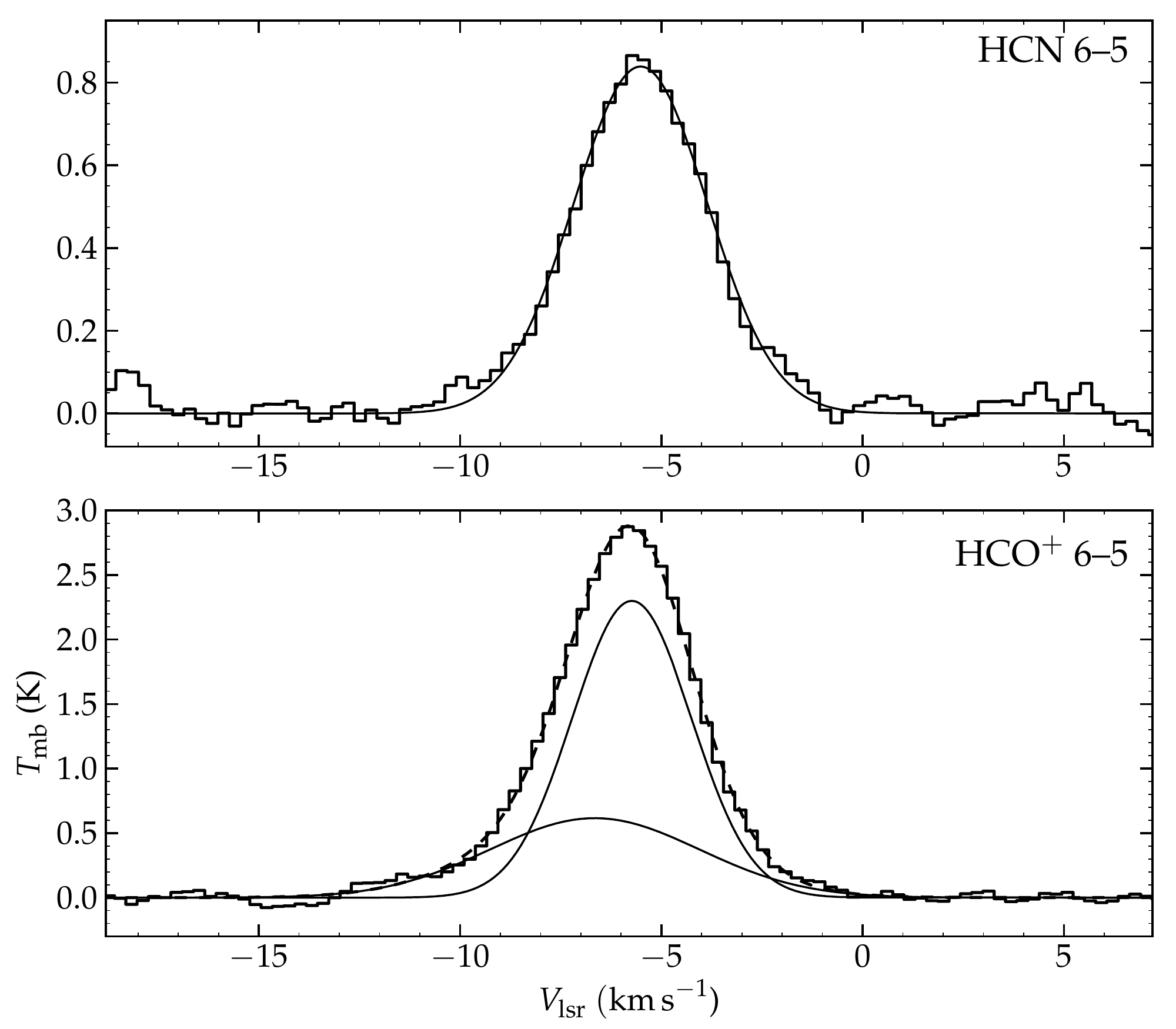}}
	 \caption{Examples of Gaussian component fits to HCN~6--5 (top, single component) and \HCOplus~6--5 (bottom, two components). Observed line profiles are shown as histograms, fit components are shown as smooth solid curves and the sum of the two components in the bottom panel is represented by the dashed curve. }
	 \label{fig:componentfits}
\end{figure}


\begin{figure*}
	\resizebox{0.5\hsize}{!}{\includegraphics{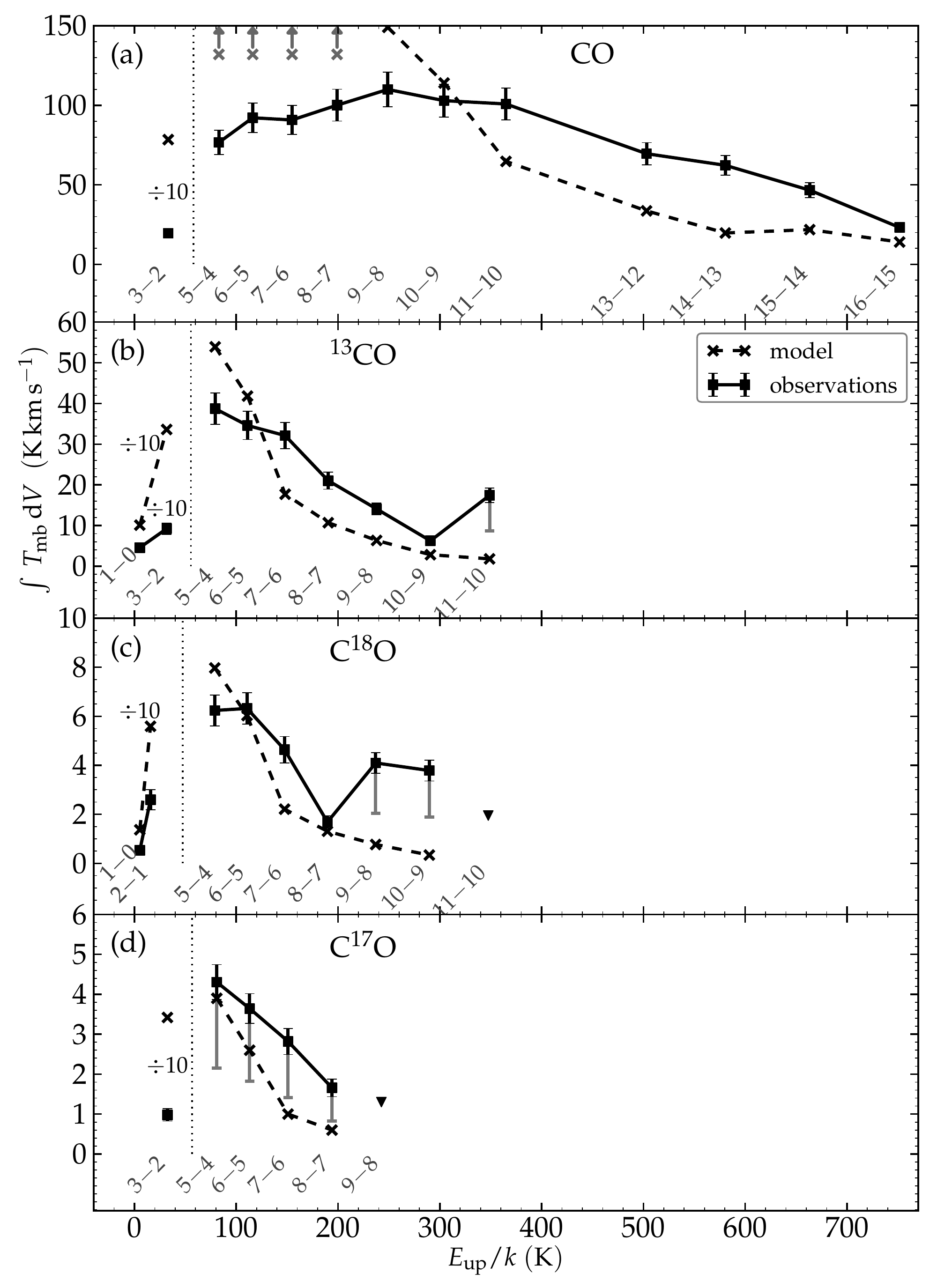}}
	\resizebox{0.5\hsize}{!}{\includegraphics{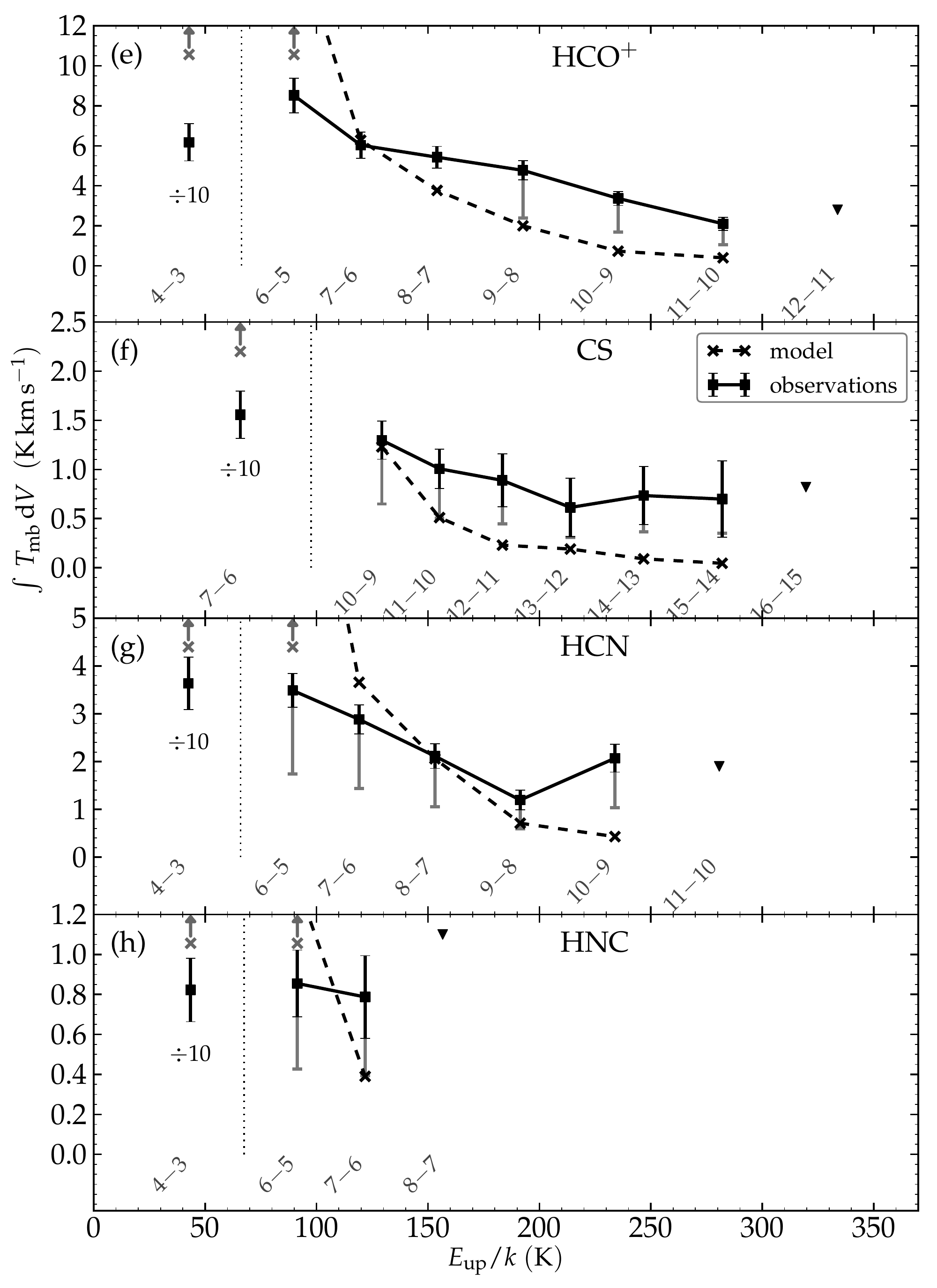}}
	 \caption{Spectral line energy distributions of CO (a), its isotopic variants (b, c, d), \HCOplus\ (e), CS (f), HCN (g) and HNC (h), as a function of \Eup. Error bars on the observations (in black) represent the quadratic sum of the absolute intensity calibration uncertainty and the formal error on the Gaussian fit. Observed intensities represent only the envelope component for those lines fitted with a double Gaussian profile (Table~\ref{t:obslines}). Since the intensity for a line fit by a single Gaussian is likely to be contaminated by outflow contribution as well, gray error bars provide an indication of the line intensity when up to half of the total line strength is attributed to the outflow. Downward pointing triangles indicate upper limits for undetected lines (see text in Sect.~\ref{sec:obsdetected}). Integrated line intensity values from the spherical model are marked by crosses and connected by dashed lines. Intensities from ground-based observations ($J$=1--0, 2--1 and 3--2 for CO species, lowest transitions shown for \HCOplus, CS, HCN and HNC) and the corresponding model points (left of the dotted vertical line) are scaled down by a factor 10, as indicated by the `$\div10$' labels. Gray crosses accompanied by upward pointing arrows denote cases where the modeled line intensity falls off the vertical scale.}
	  \label{fig:SLEDs_all}
\end{figure*}

Integrated line intensities, \intintens, are determined by fitting Gaussian profiles to the observed spectral lines (Table~\ref{t:obslines}), using the Levenberg-Marquardt fitter in the line analysis module of {\sc cassis}\footnote{{\sc cassis} is developed by CESR-UPS/CNRS (http://cassis.cesr.fr).}. First, a single Gaussian profile is fitted to each observed line. \footnote{Only in the case of CO 16--15 in band 7b is the Gaussian line fitting preceded by subtracting a linear baseline to minimize the influence from the nearby OH line in the other sideband.} In cases where this yields a residual spectrum with an rms noise comparable to or below the local baseline noise level, a single Gaussian profile is adopted. If subtraction of the single Gaussian fit leaves a significant residual signal, a fit of two blended Gaussian profiles is performed. If this gives a residual spectrum with a significantly lower rms than the one resulting from the single Gaussian fit, we adopt the double Gaussian fit. Examples of cases where a single or a double Gaussian fit is appropriate are shown in Fig.~\ref{fig:componentfits}. Although the CO lines up to $J$=5--4 may accommodate more than two Gaussian components, the additional components are so faint that they have negligible effect on the two primary components.

Lines that are well represented by a double Gaussian fit are CO, \thCO, \CeiO\ and \HCOplus\ up to a certain $J$ level, as well as all ground-based line profiles, while a single Gaussian suffices for the HIFI lines of \CstO, HCN, HNC, and CS. In general, a lower signal-to-noise of a spectral line reduces the chance of finding a meaningful secondary Gaussian component. For the lines with a double Gaussian profile, we attribute the component with the narrowest line shape and reddest \vcentr\ to the quiescent envelope, and the broader blueshifted component to outflow material. For lines with a single Gaussian profile, all emission is attributed to the envelope. For CO and \HCOplus, the component separation allows for an analysis of variations in physical conditions between the envelope and the outflow. This is addressed in Sect.~\ref{sec:outflow-column}. 

Figure~\ref{fig:SLEDs_all} shows the \intintens\ of the envelope components versus the energy of the upper level (\Eup). For the lines where only a single Gaussian fit is used, it is still likely that up to $\sim$50\% of this emission stems from the outflow component. We use a gray errorbar to indicate the intensity level that remains after 0--50\% of the single Gaussian intensity is removed from the envelope component intensity. The general trend in these spectral line energy distributions (SLEDs) is that the line intensities decrease with increasing upper level energy. A detailed comparison of the observed envelope SLEDs to models is discussed in Sect.~\ref{sec:modelSLEDs}. 

We emphasize that, although we present various trends as a function of \Eup, it is important to recognize the additional systematic effects of decreasing beam size, and decreasing line optical depth, simultaneously with increasing \Eup\ for consecutive transitions.

\subsection{Observed centroid velocities and line widths}
\label{sec:vcentr}

\begin{figure}
	\resizebox{\hsize}{!}{\includegraphics{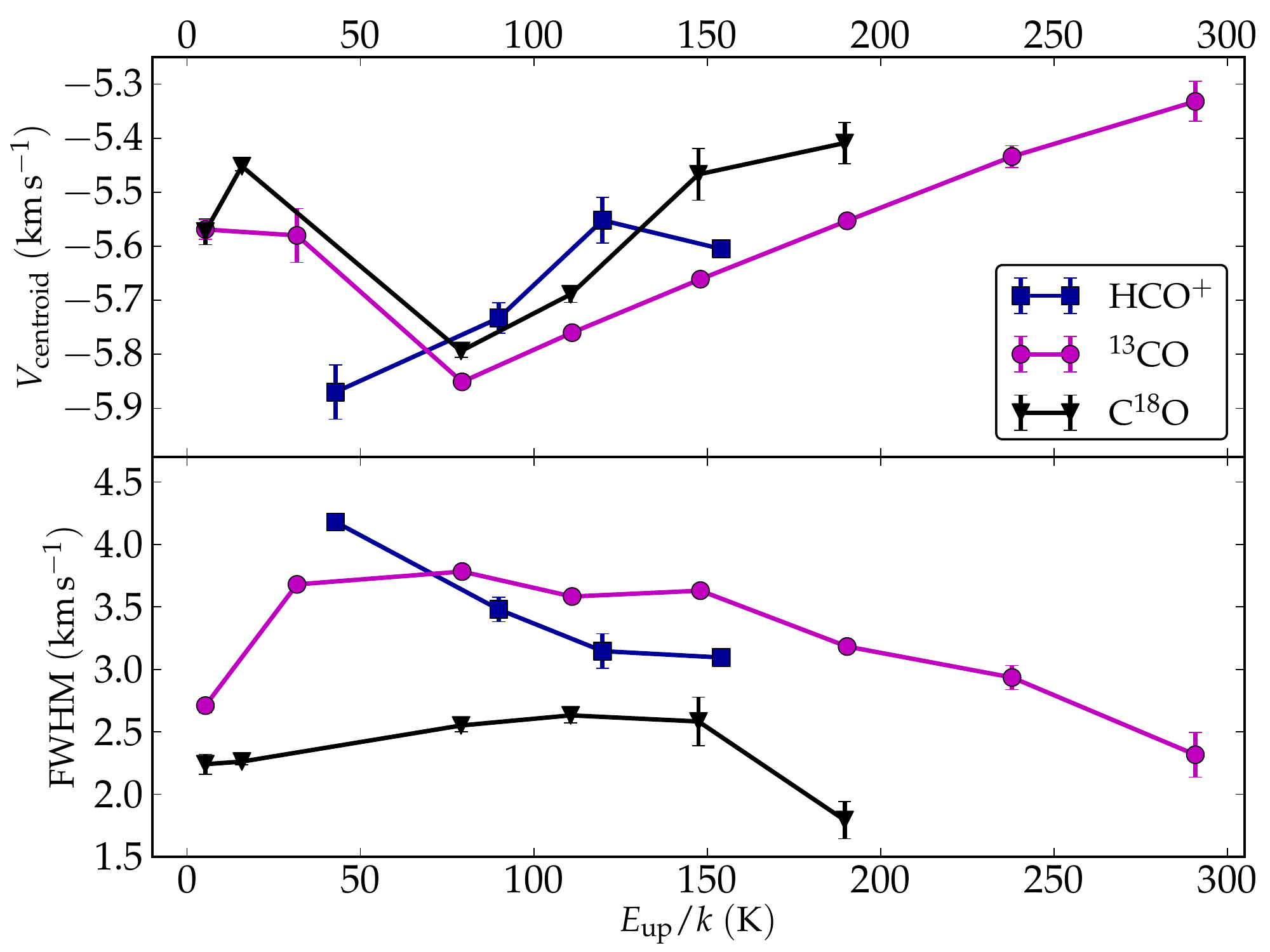}}
	 \caption{Trends of fitted centroid velocity (top) and fitted FWHM (bottom) of the envelope component versus \Eup.  }
	 \label{fig:vcentr-fwhm}
\end{figure}

From the fitted Gaussian profiles, we also obtain measures for centroid velocity and line width (Table~\ref{t:obslines}). Figure~\ref{fig:vcentr-fwhm} shows the trend of fitted values of \vcentr\ and FWHM for the envelope component, as a function of the upper level energy of the line transition. We only plot those species for which at least two lines are fitted with a double Gaussian profile. The main form of CO is excluded, since its line widths are heavily influenced by optical depth effects. 
Results from the HIFI spectral scans of the low-mass protostar IRAS16293, also from the CHESS program, show a similar trend as is found in our data of AFGL2591, but with \vcentr\ \emph{decreasing} with increasing \Eup\ (E.~Caux, private communication). We therefore conclude that the shifting centroid velocity over the HIFI bands is not an instrumental effect. 

In the diagram for \vcentr\ (Fig.~\ref{fig:vcentr-fwhm}, top), the two CO isotopologues appear to have a minimum \vcentr\ (bluest peak) at \Eup$\sim$80\,K ($J$=5--4), with a monotonously increasing trend toward higher \Eup. The four \HCOplus\ lines also exhibit an overall trend of centroid velocities becoming redder with increasing \Eup.  
To explain this systematic shift in centroid velocity, we note that increasingly higher $J$ levels do not only probe higher temperatures, parameterized by \Eup, but also progressively smaller portions of the envelope, due to the decreasing angular beam size (Tables~\ref{t:obsoverview} and \ref{t:obslines}). In cases such as the envelope of AFGL2591, which is internally heated and enjoys a radially declining temperature profile, increasing $J$ level and decreasing beam size both work in the same direction: the line signal probes progressively warmer gas. The observed velocity trend could be explained by (i) the very warmest core of the envelope being at a slightly different \vlsr\ than the surrounding envelope, or (ii) colder outer parts of the envelope being (mildly) influenced by blueshifted outflow material. The latter explanation is corroborated by the \thCO\ and \CeiO\ lines up to 3--2, which have redder \vcentr\ than their 5--4 lines observed by HIFI (Fig.~\ref{fig:CO_lineprofiles}). The lines at \Eup$<$50\,K lines are observed at IRAM~30m or JCMT with beam sizes of 22\arcsec\ (\thCO\ and \CeiO\ \mbox{1--0}), 11\arcsec\ (\CeiO~2--1) and 15\arcsec\ (\thCO~3--2 and \HCOplus~5--4), all much smaller than the 38\arcsec\ beam applicable for the lowest-$J$ HIFI lines.

\begin{figure}
	\resizebox{\hsize}{!}{\includegraphics{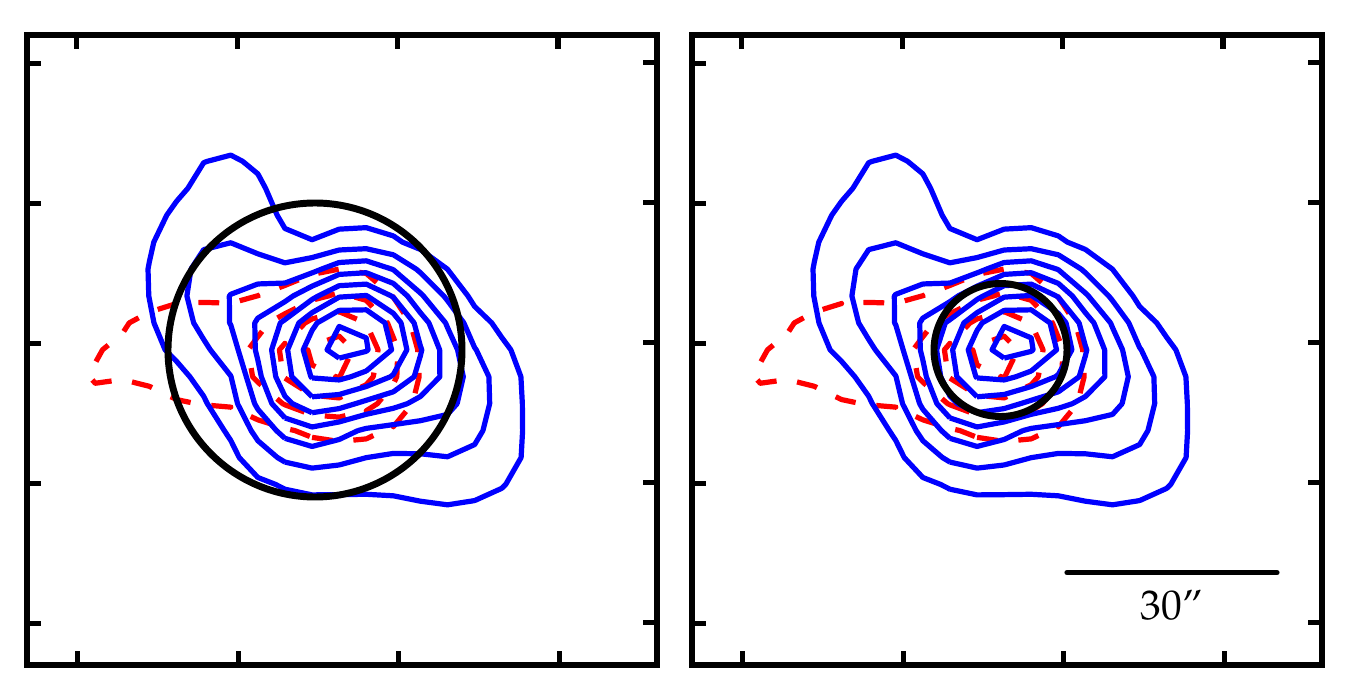}}
	 \caption{Indication of the half-power beam width of {\it Herschel}/HIFI, represented by black circles: 42\arcsec\ at 500\,GHz (left) and 19\arcsec\ at 1100\,GHz (right). The contour map represents \HCOplus~4--3 emission toward AFGL2591 measured by the HARP-B instrument at JCMT. The blue- and redshifted wings are shown in solid and dashed contours, respectively, and are defined as in \citet{vanderwiel2011}. The lowest contour and spacing between consecutive contours is 5\,\Kkms\ for both components.}
	 \label{fig:beamsize}
\end{figure}

To test the possible influence of the beam size on the centroid velocity value, we take advantage of the 1\arcmin$\times$1\arcmin\ map of \CeiO~\mbox{2--1} obtained at the IRAM 30m telescope. When the map is convolved to simulate the largest {\it Herschel} beam, the velocity centroid is measured to move from $-5.60$ to $-5.85$\,\kms, demonstrating that probing a larger portion of the envelope results in bluer line profiles. We conclude that the trend of increasing \vlsr\ with increasing $J$-level is likely to be caused by the systematically decreasing {\it Herschel} beam size (see Fig.~\ref{fig:beamsize}).

In addition, we investigate the trend of FWHM of the lines as a function of \Eup\ in Fig.~\ref{fig:vcentr-fwhm} (bottom), as measured in the envelope component of the Gaussian fits. Lines of \thCO\ and \HCOplus\ are systematically narrower with increasing \Eup\ and, again, decreasing beam size. As with the \vcentr\ trend, the two ground-based lines for \thCO\ form an exception. The systematic shift is less clear or even reversed for \CeiO.

To explain systematically decreasing line widths, a turbulent core model \citep{mckee2003} could be invoked, in which turbulent velocity increases with radial distance due to energy dissipating from the warm inner regions into the colder outer envelope. For example, \citet{herpin2012} successfully model an extensive set of \water\ lines with such a model description. If we adopt a similar model, however, line widths would be expected to increase toward lower transitions for \emph{all} species. Since the observed FWHM values for \CeiO\ between 0 and 150\,K and \thCO\ below 50\,K show no such relation, we fall back on the hypothesis that the outflow material contributes to the `envelope component' line widths. As illustrated in Fig.~\ref{fig:beamsize}, the smaller telescope beam at higher frequencies (and higher \Eup) picks up less emission from the spatially extended outflow. 

In addition, systematic trends of line velocity and line width with energy and/or beam size might be explained by an infall velocity gradient. This does not seem to be the case, however, for AFGL2591, because, even at a spectral resolution of 0.1--1\,\kms, none of our line shapes (Figs.~\ref{fig:CO_lineprofiles}, \ref{fig:nonCO_lineprofiles}) show asymmetries that could indicate infall.   

In conclusion, the trends in both \vlsr\ and FWHM suggest that the `envelope component' is kinematically affected by the outflow material, even after separating the obvious contribution of outflow material by means of the double Gaussian fits. This implies that our method to separate the outflow from the envelope is only partially successful. The assumption that the two physical components, quiescent envelope and outflow, are fully disconnected and do not influence each other dynamically obviously constitutes a highly simplified description of reality.

\subsection{Angular size of emission region}
\label{sec:HCNsize}

Information about the angular size of an emission region can be derived from direct comparison of line intensities measured at the same frequency with different telescopes and beam sizes. 
One of the few atmospheric transmission windows that allows far-infrared frequencies to be observed from the ground is the region around 800\,GHz, where this work presents the HCN~\mbox{9--8} lines at 797.433\,GHz and the CO~7--6 line at 806.652\,GHz in a 27\arcsec\ {\it Herschel} beam. These two lines were measured toward AFGL2591 by \citet{boonman2001} using the JCMT, with a beamsize of 8\arcsec\ (FWHM). The angular size of the emission region ($\theta_\mathrm{s}$) is coupled with the measured line intensities through its relation with intrinsic line brightness $T_\mathrm{s}$, observed main beam line brightness \Tmb, and telescope beam size $\theta_\mathrm{beam}$, \citep{wilson2009book}: 
\begin{equation}
T_\mathrm{s} = T_\mathrm{mb} \frac{{\theta_\mathrm{s}}^2 + {\theta_\mathrm{beam}}^2}{{\theta_\mathrm{s}}^2} ,
\label{eq:beamdilution}
\end{equation}
assuming that $\theta_\mathrm{s}$ is small compared to the telescope beam. 
Because $T_\mathrm{s}$ should be independent of the observing facility, $\theta_\mathrm{s}$ can be derived by substituting sets of (\Tmb, $\theta_\mathrm{beam}$) for both JCMT and {\it Herschel} and equating the results. 

We take the peak \Tmb\ values of CO~7--6, HCN~9--8 and HCN ($\nu_2$=$1$) 9--8 as measured at the JCMT, with an uncertainty of 50\% \citep{boonman2001}. In Table~\ref{t:sourcesize} we compare them to the peak intensities of the same lines measured in our {\it Herschel}/HIFI spectrum. For the vibrationally excited HCN, the HIFI line strength is an upper limit. Equation~(\ref{eq:beamdilution}) is then used to derive limits on the angular size of the emitting region, $\theta_\mathrm{s}$. In the limiting case of point-like emission regions in both beams, i.e., ${\theta_\mathrm{s}}^2 \ll {\theta_\mathrm{beam}}^2$, the line intensity ratio JCMT/{\it Herschel} would be $27^2/8^2\approx11$, from simple beam dilution. For larger emission regions, the intensity ratio decreases. 

The values for $\theta_\mathrm{s}$ listed in Table~\ref{t:sourcesize} imply that the HCN~\mbox{9--8} emission originates in a very confined region of the envelope of at most 1\arcsec\ in diameter. The CO~7--6 emission, however, is found to have an emission region of size $\theta_\mathrm{s} \gtrsim 10$\arcsec. Although the latter violates the original assumption that $\theta_\mathrm{s} \ll \theta_\mathrm{beam}$, it does indicate that CO~7--6 stems from a much more extended portion of the envelope than HCN~9--8. Implications of the sizes of emission regions derived here are discussed in Sect.~\ref{sec:discusssize}. 

\begin{table}
	\caption{Comparison of peak intensities of lines observed by both {\it Herschel} (3.5\,m) and JCMT (15\,m).}
	\label{t:sourcesize}
	\begin{tabular}{@{} l r r r r @{}}
	\hline
	\hline
Line transition	& \multicolumn{2}{c}{measured \Tmb\ (K)} &  \Tmb\ ratio	& inferred $\theta_\mathrm{s}$ \\
 \cline{2-3} 
			& JCMT		&  {\it Herschel}	 & & (\arcsec)	\\
	\hline 
CO 7--6				& 88$\pm$44	& 22$\pm$2		& 4$\pm$2 & $\gtrsim$10  \\
HCN 9--8				& 9$\pm$4.5	& 0.32$\pm$0.03	& 28$^{+19}_{-15}$  & $\ll$1  \\
HCN ($\nu_2$=$1$) 9--8	& 4$\pm$2	& $<$0.2			& $>$10	    & $<$3 \\
	\hline
	\end{tabular}
	\tablefoot{JCMT intensities are taken from \citet{boonman2001}, {\it Herschel} intensities are from this work. The HCN ($\nu_2$=$1$) 9--8 upper limit for {\it Herschel} is derived from the $\sim$0.1\,K 1-$\sigma$ noise level near 797.3\,GHz in our spectrum. The inferred size of the emission region, $\theta_\mathrm{s}$, is calculated using Eq.~(\ref{eq:beamdilution}). }
\end{table}

\section{Envelope spectral line energy distribution model}
\label{sec:modelSLEDs}

\subsection{Model setup}
\label{sec:modelsetup}

Radiative transfer calculations with the Monte Carlo {\sc ratran} code \citep{hogerheijde2000} are employed to provide a model comparison to the observed SLEDs from Sect.~\ref{sec:obsSLEDs}. We rely on spectroscopic parameters and collisional rates from the {\sc lamda} database \citep{schoeier2005}, which uses data provided by \citet{yang2010} for CO, \citet{dumouchel2010} for HCN and HNC, \citet{flower1999} for \HCOplus\ and \citet{turner1992} for CS. 

The physical model we use is derived from a fit to the radial intensity profile of the sectors of the 450 and 850\,\micron\ JCMT SCUBA maps that appear spherical and to the $\lambda \gtrsim 50$\,\micron\ section of the continuum spectral energy distribution (Luis Chavarr\'ia, private communication), with a strategy similar to that of \citet{vandertak1999,vandertak2000jul}. Our current model takes into account the new distance of 3.3\,kpc and a corresponding input luminosity of \pow{2.1}{5}\,$L_\odot$. Briefly, the model consists of a radial density gradient for molecular hydrogen described by an $r^{-1}$ power law (\pow{4}{7} down to \pow{7}{2}\,\pccm), a total gas mass of $\sim$500 $M_\odot$, and a corresponding temperature profile which varies monotonically from $\sim$1400\,K in the innermost shell to 27\,K in the outer shell at 70\,kAU. The radial distance at which the signal in the SCUBA 450\,\micron\ map falls below 3$\sigma$ determines the value of the outer radius of the model envelope.  
The total luminosity, as determined by integrating over the best fit model continuum spectral energy distribution, is \pow{8}{4}\,$L_\odot$; the difference with the observed luminosity is due to the absence of a cavity and a disk in our one-dimensional model, which does not include near-infrared emission.
\citet{jimenez-serra2012} have found chemical segregation at very small scales ($\lesssim$1\arcsec, 3\,kAU), but this is not expected to impact our models at the scales probed by our single dish observations ($\gtrsim$15\arcsec, 50\,kAU). Therefore, molecular abundances in our models are kept constant throughout the envelope and are scaled to match the observed SLEDs. 

After the Monte Carlo method solves for the molecular level populations, ray tracing determines the emergent line intensity, which is convolved with a Gaussian telescope beam with the appropriate HPBW at the relevant frequency. Resulting spectral maps are collapsed along the spectral direction to obtain the integrated line intensity value, which is compared to observations in Fig.~\ref{fig:SLEDs_all}. Because the outflow component is not represented in the model, we compare predicted integrated line intensities from the models described above only with the envelope component of the observed spectral lines (Sect.~\ref{sec:obsSLEDs}).

Despite the obvious increase in linear size scales when comparing an older model for AFGL2591 assuming a distance of 1\,kpc \citep[e.g.,][]{vandertak2000jul,vanderwiel2011} with the one employed in this work, the differences in emergent line intensities  are relatively minor. This is due to the similar temperature and density structure, which lead to comparable excitation conditions. A noteworthy element of the new model is the temperature of $>$1000\,K in the central shells of the envelope, compared to $372$\,K for the old model. The maximum temperature in the model is essentially set by the arbitrary choices of the thickness of the spherical model shells and the size of a central sphere which is vacant of gas. By running test models in which the very hottest inner shells are removed, we determine that the hot gas in the central envelope has no impact whatsoever on the emergent line intensities of rotational levels up to \Eup=700\,K. Essentially, none of the line emission emanating from the hot gas escapes the envelope to be detected by external observers.

\subsection{Comparison of model results with observations}
\label{sec:modelobscompare}

With a CO abundance of \pow{5}{-5} relative to \HH, one quarter of the \pow{2}{-4} applicable for warm gas \citep{vandertak1999}, the modeled SLED of the main form of CO lies a factor of 2 to 4 above the observed points up to \mbox{$J$=9--8} (off the scale in Fig.~\ref{fig:SLEDs_all}a). Beyond that, the model SLED drops off sharply and underproduces observed line intensities from \mbox{$J$=13--12} onward by factors of 2 to 3. Although with smaller discrepancies, the models for the isotopic variants \thCO, \CeiO\ and \CstO\ (with abundances of \pow{5}{-7}, \pow{6}{-8} and \pow{3}{-8}, respectively) show a similar trend in Fig.~\ref{fig:SLEDs_all}b,c,d: low-$J$ emission is over-predicted while high-$J$ emission is under-predicted. 

The \HCOplus\ model, at an abundance of \pow{1}{-8}, seems successful: it matches the envelope component of the observed lines $J$=6--5 and higher to within a factor 2 (Fig.~\ref{fig:SLEDs_all}e). However, the same model overproduces \HCOplus~4--3 by a factor 5. The models for the other species studied here, CS, HCN and HNC (abundances of 4, 6 and \pow{1}{-8}, respectively), again overproduce low-$J$ emission by factors $\sim$5 and are marginally consistent with or fall below the observed points at higher $J$ (Fig.~\ref{fig:SLEDs_all}f,g,h). 

The general picture that emerges is that our smooth, spherically symmetric models predict too much emission from cold gas ($\sim$50\,K) or too little emission from warm gas ($\gtrsim$100\,K). 
This effect persists even when we consider the possibility that the envelope contribution to the emission is overestimated by a factor of 2 for lines that are fitted by a single Gaussian (gray error bars in Fig.~\ref{fig:SLEDs_all}). For all species studied here, choosing a different molecular abundance has an effect on the vertical offset of the model SLED, especially at more optically thin transitions at high energies, but it does not sufficiently reduce the slope between \Eup$\sim$100 and 200\,K to match the observations. Our model serves to show that a spherical, homogeneous geometry offers a poor match to the true excitation balance. Therefore, the absolute molecular abundances used in this section are not meaningful by themselves. For example, the CO abundance in our model is lower than generally assumed, but still severely overproduces low-$J$ line intensities and optical depths.

A model might be considered where CO is depleted from the gas phase, to suppress specifically the low-$J$ CO emission. However, the temperature profile we use for the envelope does not reach values below 25\,K. Hence, the model gas is too warm for CO freeze-out to occur, consistent with the absence of CO ice features in mid-infrared spectra \citep{vandertak1999}. 

Our model is based on a density profile with a power law index of $\alpha=-1.0$, as outlined in Sect.~\ref{sec:modelsetup} and also used in earlier work on this source (see Sect.~\ref{sec:2591highexcintro}). We have explored {\sc ratran} models with steeper profiles ($\alpha=-1.5$), but the resulting SLEDs are very similar in shape to the ones with $\alpha=-1.0$. A steeper power law therefore does not mitigate the mismatch of the shape of the linear rotor SLEDs with the observations. 

Considering the spectral dimension, the modeled and beam-convolved line profiles of, e.g., CO and \HCOplus\ up to 7--6 and HCN up to $J$=10--9 appear broadened or even self-absorbed, indicating optically thick emission. 
For example, the optical depths at line center determined during the ray tracing, for the main isotopic form of CO, are between 90 and 10 for 3--2 up to 11--10, while only the 16--15 is marginally optically thin at 0.8. The modeled lines for the isotopologues of CO are all optically thin, as are all CS transitions except $J$=7--6. However, \HCOplus\ is optically thick until 10--9, HCN until 9--8 and HNC until 7--6. 
Indications for such high optical depths are not seen in the observed line profiles (Figs.~\ref{fig:CO_lineprofiles} and \ref{fig:nonCO_lineprofiles}). This discrepancy between observed and modeled line shapes confirms earlier suspicions \citep{vandertak1999,vanderwiel2011} that line optical depths are too high in a spherical model. 
In Sect.~\ref{sec:discusscavity}, we will suggest possible approaches to address the optical depth with more sophisticated models.

\section{Physical conditions per component}
\label{sec:componentcolumn}

\subsection{Foreground clouds}
\label{sec:column_fg}

\begin{figure}
	\resizebox{\hsize}{!}{\includegraphics{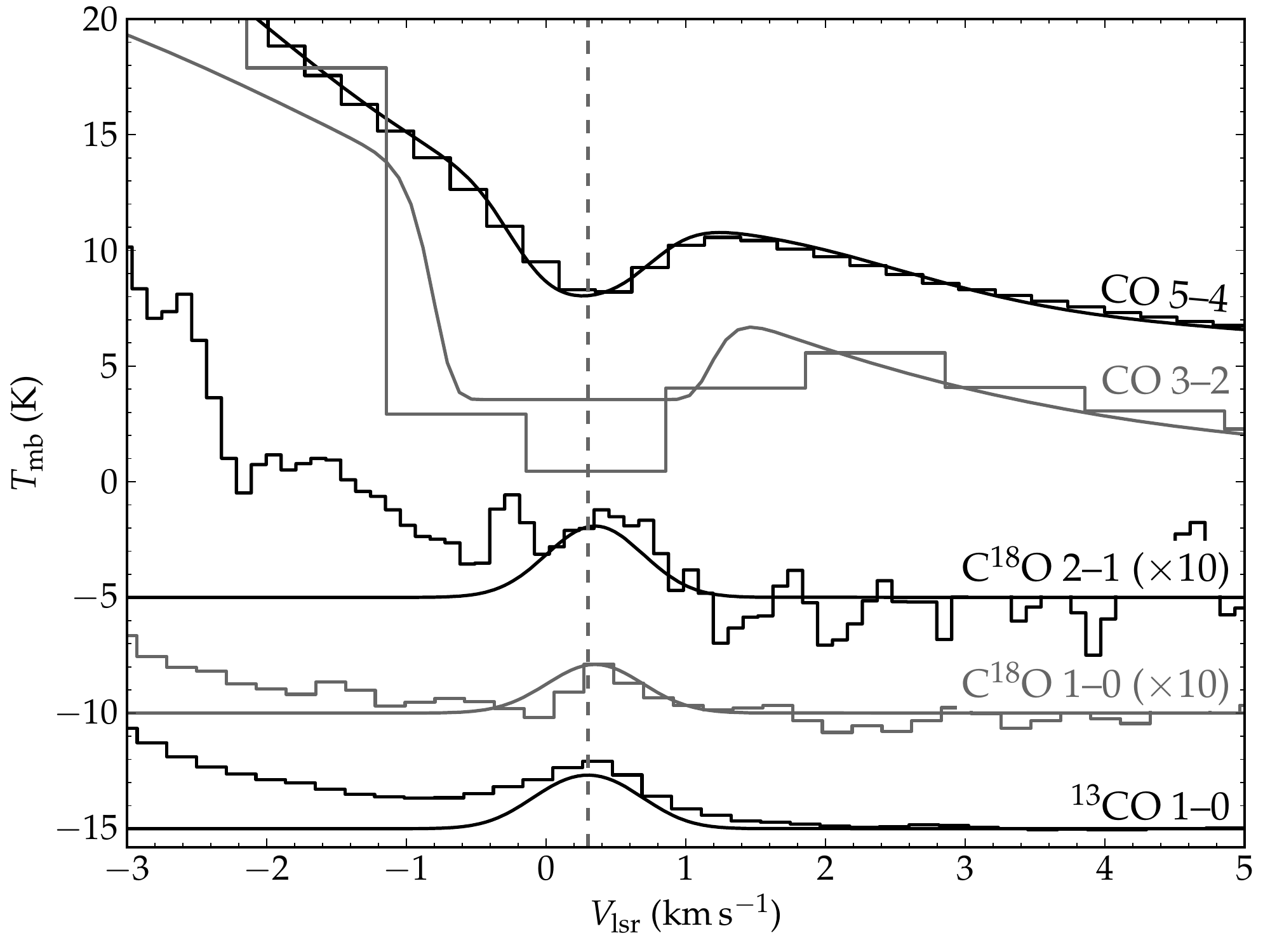}}
	 \caption{Zoom of the observed 0\,\kms~feature in various CO lines (histograms). The results from the {\sc radex} slab are shown as solid lines. To improve visibility, the brightnesses for \CeiO\ (both observations and models) are multiplied by 10. Vertical offsets are applied as follows: $+5$\,K for CO 5--4, $-5$\,K for \CeiO\ 2--1, $-10$\,K for \CeiO\ 1--0 and $-15$\,K for \thCO\ 1--0. 
	 The gray dashed line marks the approximate centroid velocity of the absorption feature at $+0.2$\,\kms. 
	 The model representations only include a component for the main envelope emission for 5--4 and 3--2, where a background is needed to absorb against. The 2--1 and 1--0 model lines only account for the foreground emission at $\sim$0\,\kms, hence the discrepancies at \vlsr~$\lesssim -1$\,\kms\ where the main envelope of AFGL2591 dominates.  }
	 \label{fig:0kmsmodel}
\end{figure}

Our CO~5--4 spectrum exhibits an absorption feature near 0\,\kms\ (Fig.~\ref{fig:0kmsmodel}), which is also seen in JCMT data of CO~\mbox{3--2} and is coincident in velocity space with a known foreground cloud \citep{vandertak1999}. Conversely, an \emph{emission} peak is visible near 0\,\kms\ in our IRAM spectra of \thCO~\mbox{1--0} and \CeiO~\mbox{1--0} (\Eup~$=5.3$\,K, Fig.~\ref{fig:0kmsmodel}). \citet{mitchell1992} show spectra of CO~2--1 toward various positions in the envelope showing the same absorption feature near \vlsr~$=0$\,\kms. It is not seen in higher-$J$ lines of CO, indicating that the absorbing foreground cloud must be relatively cold. It must also be of relatively low column density, given that the absorption feature is not present in \thCO\ and \CeiO~\mbox{5--4}. 

\citet{minh2008} observe an extended 0\,\kms\ component in CO and \thCO~1--0 emission, peaking 8\arcmin\ to the northeast and northwest of AFGL2591. Our observations show that this gas cloud also extends to the position of AFGL2591 itself, and that it must be in the foreground, since it absorbs CO~3--2 and 5--4 emission from the warm protostellar envelope. 

We construct a simple `slab' component to account for the foreground CO absorption, using the interface to {\sc radex} \citep{vandertak2007} available in {\sc cassis}. First, the shape of the CO~5--4 emission line is mimicked by three `LTE' components with excitation temperatures of 100\,K, \vcentr\ of $-4.5$, $-7.3$ and $+1.6$\,\kms, FWHM values of $5.1$, $15.7$ and $2.5$\,\kms\ and CO column densities of 6, 9 and \pow{0.3}{18}\,\psqcm. Note that these emission components are merely invoked to create representative background emission and are not meant to represent real physical conditions in the protostellar envelope. On top of the mock background emission, a {\sc radex} slab is added with $N_\mathrm{CO} = \powm{3}{17}$\,\psqcm\ and a kinetic gas temperature (\Tkin) of 10\,K. This choice for temperature is motivated by the fact that it should be lower than the minimum gas temperatures of $\sim$25\,K reached at the outer edges of the protostellar envelope of AFGL2591, while at the same time it should be high enough to result in detectable emission in the 1--0 transition of \thCO\ and \CeiO, with \Eup=5.3\,K. 

Because the foreground component is likely to be part of a larger scale cloud \citep{minh2008}, the angular size of the slab is set at 50\arcsec, larger than any of the beam sizes of our observations. Finally, the absorption component is given an intrinsic FWHM of 0.8\,\kms\ and \vlsr\ of $+0.3$\,\kms\ to match the width and position of the observed absorption feature. After the frequency-dependent optical depth calculation and the radiative transfer, the result is convolved with the relevant telescope beam to make a direct comparison to observed line profiles in Fig.~\ref{fig:0kmsmodel}.

With the parameters for the cold material chosen above to match CO~5--4, we also find self-absorption in CO~3--2 (again, against a mock background emission line) that is qualitatively consistent with the JCMT observations. Moreover, due to the 10\,K gas temperature, the same component naturally explains the emission bumps observed in \thCO~1--0, \CeiO~1--0 and even in \CeiO~2--1 (Fig.~\ref{fig:0kmsmodel}). The column densities needed for \thCO\ and \CeiO\ are \pow{3}{15} and \pow{2}{14}\,\psqcm, respectively. 

\begin{table*}[!htb]
	\caption{Physical conditions separated by component.  }
	\label{t:compcond}
	\begin{tabular}{l c c c c c c c }
	\hline
	\hline
	Component	& \vlsr & FWHM & \Tkin & \nHH & $N_\mathrm{CO}$ & $N_\mathrm{H_2}$ & $X$(\HCOplus)\tablefootmark{a} \\
				& (\kms) & (\kms) & (K) & (\pccm) & (\psqcm) 		& (\psqcm)                &  \\ 
	\hline
	Foreground A	& $+0.2$ & 0.8 & 10$^{+7}_{-1}$ & - & $\gtrsim$\pow{3}{17} & $\gtrsim$\pow{3}{21} & -  \\
	Foreground B	& $+13$  & -  & - & - & $\lesssim$\pow{6}{15} & $\sim$\pow{8}{20}--\pow{6}{21} & - \\
	Outflow & $-7$ & $\sim$6--10 & 60--200 & $\sim$$10^5$--$10^6$ & \pow{1.5}{18}\tablefootmark{b} & \pow{1.5}{22} & $\sim$$10^{-9}$--$10^{-8}$  \\
	Envelope\tablefootmark{c} & $-5.7$ & $\sim$3--4 & 23--372 & $\sim$$10^3$--\pow{4}{7} & $\sim$$10^{18}$ & $\sim$$10^{22}$ & \pow{4}{-10}--\pow{1}{-8}  \\
	\hline
	\end{tabular}
	\tablefoot{A `-' symbol indicates that the corresponding quantity is unknown for that component.}
	\tablefoottext{a}{Molecular abundance of \HCOplus\ relative to \HH.} \\
	\tablefoottext{b}{Scaled from $N_\mathrm{C^{18}O}=\powm{3}{15}$\,\psqcm. } \\
	\tablefoottext{c}{Envelope number density and kinetic temperature ranges are taken from the parameterization of the spherically symmetric model (Sect.~\ref{sec:modelSLEDs}).}
\end{table*}

We conclude that a cold slab of foreground gas explains the 0\,\kms\ feature. 
Assuming an abundance of CO/\HH\ of at most $10^{-4}$, we derive a column density $N_\mathrm{H_2} \gtrsim \powm{3}{21}$\,\psqcm. The visual extinction ($A_V$) for this column density would be $\sim$3 magnitudes, following \citet{bohlin1978}, purely for the molecular part. If we add to this 1 magnitude each (front and back) for the atomic surfaces of the cloud where CO is photodissociated, we find that the total $A_V \gtrsim 5$.
The derived conditions in the 0\,\kms\ component (`foreground A') are summarized in Table~\ref{t:compcond}. 

Whereas the 1--0 emission features are insensitive to changes in \Tkin\ between 5 and 20\,K, the depth and shape of the CO~5--4 absorption features constrains \Tkin\ to [9,15]\,K. The value adopted for $N_\mathrm{CO}$ is motivated by the depth of the self-absorption feature, which starts to deviate significantly when $N_\mathrm{CO}$ is changed by a factor $\gtrsim2$. The model result is insensitive to the number density of the collision partner in {\sc radex} as long as $n$(\HH) is above $\sim$$10^4$\,\pccm, i.e., when the low-$J$ CO excitation balance is in LTE. 

Given the inferred nature of the foreground cloud, it is important to consider that \HH\ densities in translucent clouds are generally in the order of $10^3$ rather than $>$$10^4$\,\pccm. With a lower \HH\ density, the $J$=1--0 emission features in \thCO\ and \CeiO\ can be explained with temperatures as high as 17\,K. Hence, the density as well as the upper limit to the kinetic temperature of the foreground cloud are poorly constrained. 

In addition, a discrepancy is found with respect to the observed CO~6--5 line profile that shows no absorption at 0\,\kms. In fact, if any feature is present at that velocity, it is in emission (Fig.~\ref{fig:CO_lineprofiles}). The slab model with \Tkin~=~10\,K and $N_\mathrm{CO} =  \powm{3}{17}$\,\psqcm, however, predicts a detectable absorption feature for CO~6--5. We are unable to explain the CO~6--5 without compromising the match to the lower-$J$ 0\,\kms\ feature. 

Although not included in the considerations in this paper, several additional probes of the 0\,\kms\ foreground cloud are available. Firstly, besides the CO signatures shown in Fig.~\ref{fig:0kmsmodel}, the $^3$P$_1$--$^3$P$_0$ line of neutral carbon (at 492.161\,GHz in our CHESS spectral scan) also shows a weak emission feature near 0\,\kms. Secondly, an upper limit on the density might be derived from the fact that the 0\,\kms\ component is not seen in lines of, e.g., \HCOplus\ and HCN (Fig.~\ref{fig:nonCO_lineprofiles}). Thirdly, signatures of CH, CH$^+$, H$_2$O$^+$ and other hydride species do show absorption near 0\,\kms\ \citep{bruderer2010b}, as does the HF signature in the CHESS spectrum detected by \citet{emprechtinger2012}. A more extensive exploration incorporating all observed tracers of the 0\,\kms\ component toward AFGL2591 is therefore warranted.


In addition to the 0\,\kms\ component described above, there is another absorption component, at $+13$\,\kms, visible in {\it Herschel} observations of HF \citep[][but not discussed there]{emprechtinger2012} as well as \water\ \citep{choi_inprep}. It is not seen, however, in any lines of CO species presented in this work, neither those observed with HIFI (Fig.~\ref{fig:CO_lineprofiles}) nor those observed at IRAM\,30m and JCMT (the velocity range covered in the observations is larger than what is shown in Fig.~\ref{fig:0kmsmodel}). The diatomic hydrides observed in the {\it Herschel} WISH program \citep{bruderer2010b} do not show clear evidence of emission or absorption at $+13$\,\kms\ either. 

Based on our \thCO~1--0 spectrum, the $+13$\,\kms\ feature would need to have a peak intensity of $\lesssim$0.1\,K to ensure a non-detection. With that upper limit, a brief exploration of {\sc radex} models tells us that the \thCO\ column density is at most $10^{14}$\,\psqcm. Using the canonical scaling factors to convert to CO (60) and noting that the CO abundance relative to \HH\ can be as low as $10^{-6}$ in diffuse gas, we derive that $N_\mathrm{H_2}$~$\lesssim$~\pow{6}{21}\,\psqcm\ in the $+13$\,\kms\ component. 
In addition, the CHESS spectrum of HF~\mbox{1--0} presented by \citet{emprechtinger2012} provides a \emph{lower} limit for the column density of the $+13$\,\kms\ cloud. While \citeauthor{emprechtinger2012} do not discuss the $+13$\,\kms\ component, an absorption feature does appear at that velocity with a depth of $\sim$70\% relative to the saturated 0\,\kms\ component. For the latter, they derive $N_\mathrm{HF}$ $>$ \pow{4}{13}\,\psqcm. Scaling by the relative depth of the absorption features and assuming that all fluorine is in the form of HF (HF/\HH=\pow{3.6}{-8}), we find that $N_\mathrm{H_2}$ $\gtrsim$ \pow{8}{20}\,\psqcm. Column density limits are summarized in Table~\ref{t:compcond} (`foreground B').

\subsection{Outflow gas}
\label{sec:outflow-column}

Previous studies indicate that CO species, \HCOplus, and neutral atomic carbon are present in the outflow in significant amounts \citep[e.g.,][]{choi1994,hasegawa1995,doty2002}. Our results confirm that line profiles of CO species and of \HCOplus\ are well represented by two separate kinematic components (Sect.~\ref{sec:obsSLEDs}). 
 
\begin{figure}
	\resizebox{\hsize}{!}{\includegraphics{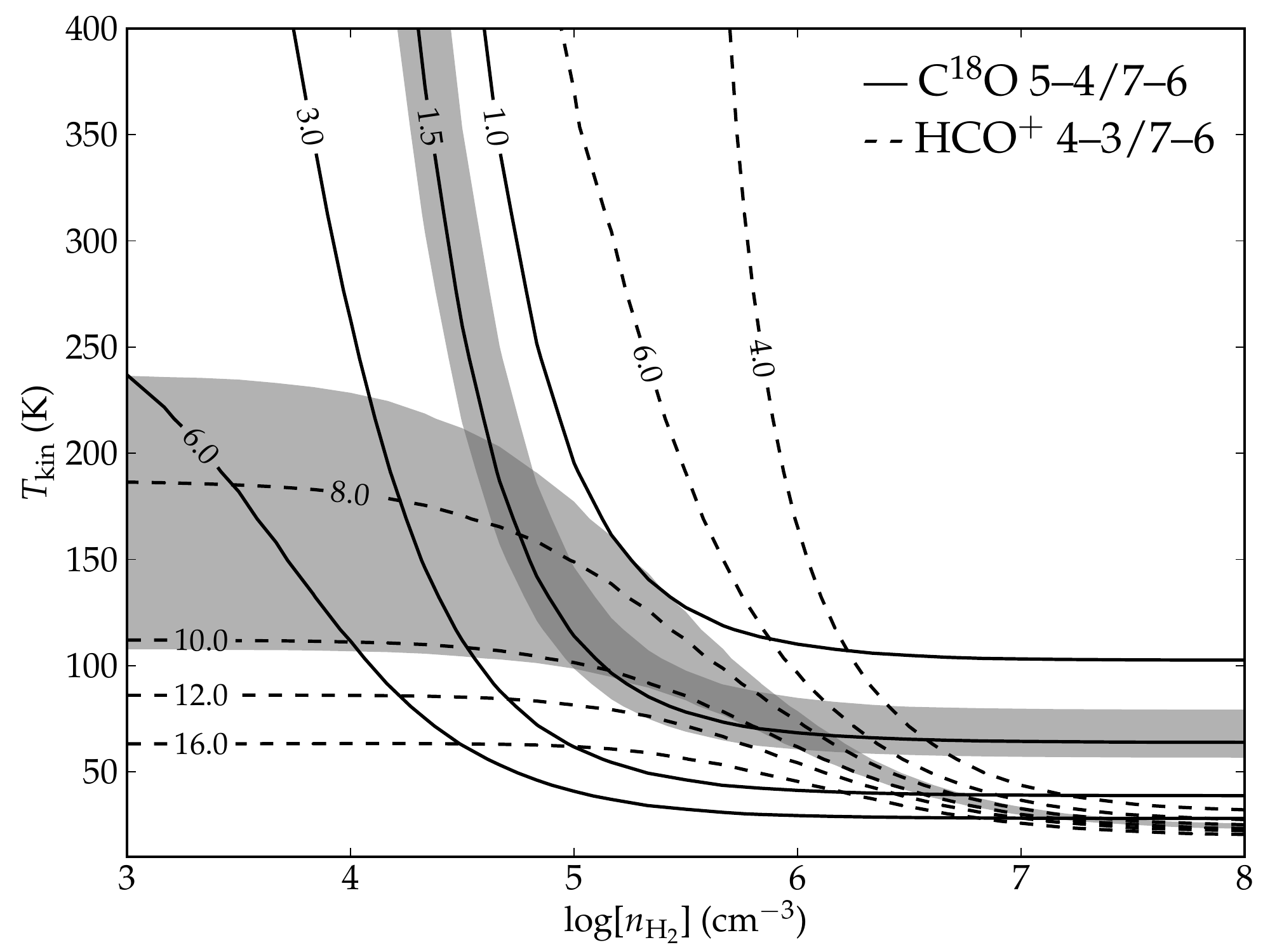}}
	 \caption{Lines of constant intensity ratio of \CeiO~\mbox{5--4/7--6} (solid lines) and \HCOplus~\mbox{4--3/7--6} (dashed lines), resulting from a {\sc radex} grid calculation. Curves are labeled by their line ratio value. The gray areas mark the 1-$\sigma$ uncertainty ranges of the observed line intensity ratios in the outflow component, including the fit error (see~Table~\ref{t:obslines}) and the 10\% calibration uncertainty (see Sect.~\ref{sec:hifiobs}). The darker gray area denotes the region where the observed line intensity ratios of both tracers overlap. }
	  \label{fig:lineratios}
\end{figure}

Using radiative transfer calculations with the {\sc radex} code, we estimate kinetic temperatures, number densities and column densities of CO and \HCOplus\ in the outflow component. This approach requires an assumption of an isothermal homogeneous material, in which the radiative transfer is solved using collisional rates from the LAMDA database \citep{schoeier2005}. 
Following \citet{vandertak2007}, we construct grids of {\sc radex} models and calculate line intensity ratios for values of \nHH\ between $10^3$ and $10^8$\,\pccm\ and \Tkin\ between 10 and 400\,K. In the input for the grid calculation, the column densities are intentionally low in order to keep all modeled lines optically thin. In this limit, the intensity of each line is directly proportional to the column density and line \emph{ratios} are therefore insensitive to the absolute column density. In Fig.~\ref{fig:lineratios} we show the resulting line ratios of \HCOplus~\mbox{4--3/7--6} and \CeiO~\mbox{5--4/7--6} as a function of \nHH\ and \Tkin. When the observed line intensity ratio is indicated in the diagram, it is possible to constrain a region in parameter space that fits the observations. 

First, the mere presence of high-$J$ \HCOplus\ emission in the outflow component is an indication that number densities must be relatively high, given that critical densities are in the order of $10^7$\,\pccm\ for the 6--5, 7--6 and 8--7 transitions. Second, from the \CeiO\ line ratio in Fig.~\ref{fig:lineratios}, it is evident that densities below $10^4$\,\pccm\ require temperatures that are unphysically high. The observed \HCOplus\ and \CeiO\ line ratios intersect in a region in (\nHH,\Tkin)-space near $10^{4.7}$--$10^6$\,\pccm\ and $\sim$60--200\,K. Inside this region, higher densities imply lower temperatures, and vice versa (Fig.~\ref{fig:lineratios}). The measured line ratio for \thCO~\mbox{5--4/7--6} gives the same constraints as \CeiO, confirming that signal-to-noise and optical depth do not affect our result. Finally, a line ratio diagram for \CeiO~\mbox{1--0/2--1} suggests the presence of an additional component of colder gas at \Tkin\ $<$ 50\,K. However, this result should be considered with caution, because (i) the kinematic separation of envelope and outflow contributions is not unambiguous (Sect.~\ref{sec:obsSLEDs}), and (ii) the beam filling factor of the outflow component could be different for the 2--1 and 1--0 observations (beam HPBW 11\arcsec\ and 22\arcsec, respectively). 

With \nHH=[$10^5$,$10^6$]\,\psqcm\ and \Tkin=[60,200]\,K, a column density in the order of $10^{13}$ to $10^{14}$\,\psqcm\ for \HCOplus\ reproduces the line strengths observed for the outflow. Similarly, a \CeiO\ column density of \pow{3}{15}\,\psqcm\ ($\pm 30\%$) is required to explain the HIFI \CeiO\ line strengths in the outflow component. With a CO/\CeiO\ ratio of 500 and a CO abundance of $10^{-4}$ relative to \HH, we derive $N_\mathrm{H_2}$~=~\pow{1.5}{22}\,\psqcm\ and a molecular abundance in the order of $10^{-9}$ to $10^{-8}$ for \HCOplus\ in the outflow gas. This \HCOplus\ abundance falls in the same range as previously adopted values for the envelope, ranging from \pow{4}{-10} \citep{carr1995} to $\sim$\pow{1}{-8} \citep{vanderwiel2011}. 

In conclusion, Table~\ref{t:compcond} contrasts the physical conditions derived here for the protostellar outflow component (Sect.~\ref{sec:outflow-column}) with those found for the CO foreground cloud (Sect.~\ref{sec:column_fg}) and in the envelope. Our current analysis indicates that the outflow gas is not significantly different from that in the envelope, considering gas density, gas temperature, as well as the chemical balance of CO and \HCOplus.

\section{Discussion and conclusions}
\label{sec:discussion}

\subsection{Envelope structure}
\label{sec:discusscavity}

The spherically symmetric radiative transfer models are unable to simultaneously explain the low- and high-$J$ transitions of the observed integrated line intensities (Sect.~\ref{sec:modelSLEDs}, Fig.~\ref{fig:SLEDs_all}): the modeled SLEDs of all species studied here decline too steeply between $\sim$100 and 200\,K. 
This indicates that the passively heated, spherical envelope structure does not populate the high $J$ levels sufficiently, or that emission from those levels is too optically thick. Line optical depths are known to be higher than observed for AFGL2591 in ground-based low excitation lines \citep{vanderwiel2011} and in Sect.~\ref{sec:modelSLEDs} we have noted that optical depths for the modeled higher $J$ lines are lower, but still $>$1 in many cases. 

To boost line intensities at \Eup\ above $\sim$150\,K without increasing the lower level intensities too much, it would be appropriate to apply a model geometry that includes outflow cavities. This would affect not only the density structure to create lower opacity pathways (if a suitable inclination angle is chosen), but also the chemical and thermal balance through direct illumination of envelope material by UV radiation from the central massive protostar. Indeed, we demonstrate in Sect.~\ref{sec:modelSLEDs} that the hot gas ($>$1000\,K) in the center of the model envelope does not contribute to emergent line intensity if it is embedded in an isotropic envelope. 

In this light, our AFGL2591 observations should be compared with a model scheme such as that by \citet{bruderer2009b,bruderer2010a,bruderer2012}, with a physical geometry appropriate for the massive protostellar envelope of AFGL2591. Their model allows for an outflow cavity to be present in an axisymmetric (or even three-dimensional) density distribution, from which the dust temperature profile is derived through continuum radiative transfer. The UV field is calculated and used as input to iteratively determine the position-dependent chemical balance, excitation balance, gas temperature and resulting line emission. The fully self-consistent chemical treatment includes freeze-out, photodissociation, X-ray ionization and other processes. Preliminary exploration of similar models already shows that, as the cavity allows a larger volume of lower density gas to be irradiated and heated directly, SLED shapes become significantly flatter compared with spherically symmetric geometries (\citealt[][Chapter~4]{vanderwiel2011thesis}; Steven~Doty 2011, private communication). \citet{bruderer2012} also show that in exposed layers of the gas envelope, the gas and dust temperature become decoupled, rendering inadequate our model for which the gas temperature is derived from dust continuum measurements (Sect.~\ref{sec:modelSLEDs}). 

The \HCOplus\ molecule in particular could provide an interesting probe of outflow cavity geometry, since, contrary to CO and HCN, the \HCOplus\ abundance is expected to be enhanced in the cavity walls (Simon~Bruderer 2011, private communication). 
Additional constraints could come from the incorporation of CO$^+$ observations. Although no signatures from this species are detected in the spectral scans with {\it Herschel}/HIFI \citepalias{kazmierczak_inprep} and with JCMT/HARP-B \citep[][Sect.~3.A]{vanderwiel2011thesis}, two submillimeter CO$^+$ lines are detected in deeper spectra by \citet{stauber2007}. From the sample of low-mass and high-mass protostars in their study, they conclude that X-rays play a role in the dissociation and excitation of molecules in low-mass protostellar envelopes, but that CO$^+$ observations can be explained by far-UV radiation in high-mass objects. Even in dissociation regions, CO$^+$ emission could be weak or absent, since this ion is easily destroyed by collisions \citep{stauber2009}. 

Additionally, line opacity properties of the modeled protostellar envelope can be adjusted by distributing the same amount of gaseous material inhomogeneously, as previously suggested by \citet{vanderwiel2011}. This could be used to further  alleviate the optical depth effects (Sect.~\ref{sec:modelSLEDs}), if the introduction of cavities in a two-dimensional geometry does not prove sufficiently effective to match the observations. 

Studies of low-mass star-forming envelopes have shown that a passively heated envelope is the dominant component for lines with \Eup~$\lesssim$~200\,K, but that the $\sim$200--4000\,K section of the CO SLED is a discriminating probe of the contribution of UV irradiation and shocks \citep{vankempen2010a,visser2012}. 
While the highest $J$ CO line presented in this work is \mbox{16--15}, the highest attainable to HIFI, {\it Herschel}/PACS has observed AFGL2591 at frequencies covering lines up to $J_\mathrm{up} \approx 30$, to be analyzed in the context of the WISH key program \citep{vandishoeck2011}. We expect the extension of the CO ladder to higher energies to shed more light on excitation drivers in massive protostellar envelopes such as AFGL2591.




\subsection{Foreground clouds}
\label{sec:discuss_foreground}

We identify a 0\,\kms\ foreground cloud in Sect.~\ref{sec:column_fg}, for which we derive a kinetic gas temperature between 9 and $\sim$17\,K, based on the combination of absorption and emission signatures in the $J$=1--0 up to 5--4 transitions of CO and isotopologues. 
The material in the foreground cloud discussed here is redshifted with respect to AFGL2591. Therefore, the foreground could be feeding additional gas to the protostellar system, but evidence would have to be found to support physical proximity of the 0\,\kms\ cloud to AFGL2591. 

For a second foreground cloud, at \vlsr=$+13$\,\kms, we use our non-detections in CO and the absorption in HF to derive limits for the column density: \pow{8}{20}\,\psqcm\ $\lesssim$ $N_\mathrm{H_2}$ $\lesssim$ \pow{6}{21}\,\psqcm\ (Sect.~\ref{sec:column_fg}). The CO abundance relative to \HH\ must be $\lesssim10^{-5}$ for the derived limits to be consistent with each other (for CO/\HH=$10^{-4}$, the upper limit would be \pow{6}{19}\,\psqcm). 
In such diffuse conditions HF is more abundant than CO, as predicted by \citet{neufeld2005} and observed, for example, by \citet{sonnentrucker2010} on the line of sight to W51. It is surprising, however, to detect a relatively strong absorption signature of \water\ without the presence of CO. Water abundance values will be derived by \citet{choi_inprep}.

\subsection{Outflow gas}
\label{sec:discuss_outflow}

The kinetic temperature of 60--200\,K and number density of $\sim$$10^5$--$10^6$\,\pccm\ derived for the outflow gas are comparable to those in the passive envelope (Sect.~\ref{sec:outflow-column}). The component that we identify a few \kms\ blueward of the systemic \vlsr\ is kinematically distinct from the envelope (Table~\ref{t:obslines}), but it apparently harbors gas that is similar in density and temperature. Our results are consistent with conditions found in other protostellar outflows \citep[e.g.,][]{giannini2001a,moro-martin2001b,lefloch2010,yildiz2010,yildiz2012,bjerkeli2011,gomez-ruiz2012}. Specifically, CHESS spectra toward the L1157-B1 outflow reveal \HH\ gas densities of $\sim$$10^5$ to a few $10^6$\,\pccm\ and temperatures around 200\,K \citep{codella2012a,benedettini2012,lefloch2010,lefloch2012}.

We speculate that the similar physical and chemical conditions (Table~\ref{t:compcond}) between outflow and envelope could point to outflow gas that has been expelled from the envelope relatively recently. The blue component observed in this work may in fact trace swept-up envelope material ($\lesssim$10\,\kms\ from the systemic velocity) rather than fast outflowing gas that is farther separated from the parent envelope, see, e.g., \citet{vandertak1999} for AFGL2591, or \citet{qiu2011} and \citet{gomez-ruiz2012} for other protostars. Our outflow conditions are compatible with the temperature and density derived for the warmest of the three outflow components in L1157-B1, as found by \citet{lefloch2012}. These authors argue that lower kinetic temperatures indicate older outflows, analogous to our conclusion that the warm, dense outflow material from AFGL2591 is young. 
We should also caution that the double Gaussian fits that are applied to distinguish the components (Sect.~\ref{sec:obsSLEDs}) are likely to suffer from cross-contamination. Both components could partly trace the same gas and, as a consequence, the derived physical conditions for the envelope and outflow may be artificially equalized.

\subsection{Compact HCN emission}
\label{sec:discusssize}

In Sect.~\ref{sec:HCNsize}, our observations of HCN~9--8 and CO~7--6 are compared with those from a larger telescope, in order to impose limits on the size of the emitting regions. We find sizes of $\lesssim$$1\arcsec$ for HCN~9--8 and $\gtrsim$10\arcsec\ for  CO~\mbox{7--6}. 
\citet{benz2007} find source sizes of $<$1\arcsec\ in an interferometric HCN~4--3 map, but the extended emission in their observation is likely filtered out. Conversely, single dish observations by \citet{vanderwiel2011} reveal larger scale ($>$$ 30\arcsec$) emission in the 4--3 transition of HCN, but this line traces lower temperatures and lower densities than HCN~9--8. Considering our envelope model from Sect.~\ref{sec:modelSLEDs}, only in the very inner shells ($<$ 30\,AU $\approx$ 0.01\arcsec) does the number density come within an order of magnitude of the critical density of HCN~\mbox{9--8}, \pow{3}{8}\,\pccm. It is therefore conceivable, and fully consistent with our observations, that the HCN $J$=9--8 emission arises in a region smaller than 1\arcsec\ in angular size. The critical density for CO~7--6 is a factor $\sim$50 lower than that of HCN~9--8; it is therefore expected that CO~7--6 emerges from a more extended region of the envelope.

\section{Summary and outlook}
\label{sec:summaryoutlook}

In this first {\it Herschel} CHESS paper on the massive star-forming envelope AFGL2591, we present HIFI spectral line measurements of a wide range of transitions (\Eup/$k$=80 up to 300--700\,K) of CO, \thCO, \CeiO, \CstO,  \HCOplus, CS, HCN and HNC, supported by ground-based measurement of lower excitation levels. Emission lines are separated into a narrow `envelope' and a broader `outflow' component, if the $S/N$ is sufficient (Sect.~\ref{sec:obsSLEDs}). The integrated line intensities of the envelope component, presented in the form of spectral line energy distributions (SLEDs), are compared with radiative transfer results based on a spherically symmetric envelope model (Sect.~\ref{sec:modelSLEDs}).  

The main conclusions of this work are the following. 
\begin{enumerate}
	\item The model SLEDs for all species fall off too steeply between 100 and 200\,K to explain the flat shape of the observed envelope SLEDs (Sect.~\ref{sec:modelSLEDs}). We expect that a model including outflow cavities and direct UV illumination will provide a better match (Sect.~\ref{sec:discusscavity}).
	\item The outflow gas is found to be as dense and warm as that in the quiescent envelope (Sect.~\ref{sec:outflow-column}). We conclude that the outflow gas traced here may have been expelled only recently (Sect.~\ref{sec:discuss_outflow}).
	\item Trends of line centroid and line width with frequency, excitation level or telescope beam size are identified in the envelope component. In absence of a single, consistent, physical explanation, we ascribe these trends to ambiguities in the separation of the envelope and outflow components (Sect.~\ref{sec:vcentr}). 
	\item By comparing line intensities measured with {\it Herschel} (3.5\,m) and JCMT (15\,m), we conclude that HCN~\mbox{9--8} emission must originate from a region $<$1\arcsec\ in angular size (Sects.~\ref{sec:HCNsize} and \ref{sec:discusssize}).
	\item A known foreground cloud at \vlsr=0\,\kms\ is detected in several lines of CO and isotopologues. Limits to its temperature and column density are derived in Sect.~\ref{sec:column_fg}.
	\item A second foreground cloud, detected in HF and \water\ at $+13$\,\kms, remains undetected in our observations (Sect.~\ref{sec:column_fg}). Its nature and column density are discussed in Sect.~\ref{sec:discuss_foreground}.
\end{enumerate}

Future papers in this series based on the CHESS observations of AFGL2591 will be ordered by chemical families of species, e.g., N-bearing species \citepalias{kazmierczak_inprep}, S-bearing species. \citetalias{kazmierczak_inprep} will also include details of all other molecular line signal detected in the 490--1240\,GHz spectral scan. 
In addition, future work should be devoted to the application of a more sophisticated model geometry for AFGL2591 to explain the highly excited rotational lines accessible with {\it Herschel}. Such a model should include at the least an outflow cavity geometry and chemical effects of direct UV illumination (now available for hydrides and C/C$^+$/CO, \citealt{bruderer2010a,bruderer2010b,bruderer2012}), but possibly also shock physics and inhomogeneous envelope structure.

\begin{acknowledgements}
The authors acknowledge constructive discussions with members of the HIFI Instrument Control Center. We thank Mihkel Kama and the other members of the CHESS data reduction team for frequent interaction on HIFI spectral scan data processing. 
We thank Luis Chavarr\'ia for providing the new spherical envelope model structure, 
and Marco Spaans, Steve Doty, Ren\'e Plume and Simon Bruderer for discussions on physical and chemical modeling. 
We acknowledge constructive comments by an anonymous referee and by the Editor. \\
HIFI has been designed and built by a consortium of institutes and university departments from across Europe, Canada and the United States under the leadership of SRON Netherlands Institute for Space Research, Groningen, The Netherlands and with major contributions from Germany, France and the US. Consortium members are: Canada: CSA, U.\,Waterloo; France: CESR, LAB, LERMA, IRAM; Germany: KOSMA, MPIfR, MPS; Ireland, NUI Maynooth; Italy: ASI, IFSI-INAF, Osservatorio Astrofisico di Arcetri-INAF; Netherlands: SRON, TUD; Poland: CAMK, CBK; Spain: Observatorio Astron\'omico Nacional (IGN), Centro de Astrobiolog\'ia (CSIC-INTA). Sweden: Chalmers University of Technology - MC2, RSS \& GARD; Onsala Space Observatory; Swedish National Space Board, Stockholm University - Stockholm Observatory; Switzerland: ETH Z\"urich, FHNW; USA: Caltech, JPL, NHSC. \\
This work is supported in part by the Canadian Space Agency and NSERC. 

\end{acknowledgements}

\bibliographystyle{bibtex/aa}  
\bibliography{../../literature/allreferences}

\end{document}